\documentclass[a4paper,10pt]{article}
\usepackage{amsmath}
\usepackage{mathrsfs}
\usepackage{bm}
\usepackage{fullpage}
\usepackage{enumerate}
\usepackage{hyperref}
\newcommand{\upd}{\mathrm{d}}
\allowdisplaybreaks[1]
\begin{document}

\title{Stability, Instability, Canonical Energy and Charged Black Holes}
\author{Joe Keir \\ \\
{\small DAMTP, Centre for Mathematical Sciences, University of Cambridge}, \\
{\small \sl Wilberforce Road, Cambridge CB3 0WA, UK} \\ \\
\small{j.keir@damtp.cam.ac.uk}}
\maketitle

\begin{abstract}
We use the \emph{canonical energy method} of Hollands and Wald to study the stability properties of asymptotically flat, stationary solutions to a very general class of theories, consisting of a set of coupled scalar fields and $p$-form gauge fields, minimally coupled to gravity. We find that, provided certain very weak assumptions are made on the coupling coefficients, the canonical energy method can be extended to this class of theories. In particular, we construct a quadratic form $\mathcal{E}$ on initial data perturbations, with the properties that $\mathcal{E} > 0$ on all perturbations indicates stability, while $\mathcal{E} < 0$ on some perturbation indicates instability. Furthermore, we show that the conditions needed for the existence of $\mathcal{E}$ allow for a stable definition of asymptotic flatness. Finally, we extend the proof of the Gubser-Mitra conjecture, given by Hollands and Wald, to this class of theories. In particular, this shows that for sufficiently extended, charged black brane solutions to such theories, thermodynamic instability implies dynamical instability.
\end{abstract}

\section{Introduction}
The stability properties of solutions to Einstein's equations in various dimensions and in the presence of various matter fields are of considerable interest, both in themselves and in the context of low energy approximations to string theory. However, even in the vacuum, asymptotically flat, four dimensional case relatively little is known about stability beyond the linear level. In particular, the Kerr solution has not been proved to be stable, although both linearised analytic calculations and, numerical calculations indicate that it is \cite{Dafermos:2010hb} \cite{Whiting:1988vc} \cite{Krivan:1997hc}.

In contrast to the situation in four dimensions (in which stationary solutions are believed to be stable), recent work (see \cite{Emparan:2003sy}) has shown that higher dimensional, stationary vacuum solutions are not all stable. In particular, black holes which rotate fast enough have been shown to be unstable. The landscape of ``black objects'' is also more complicated, including black rings \cite{Emparan:2001wn} and possibly other, more complicated objects. When matter fields are introduced, the situation can become even more complicated.

Another class of solutions exhibits interesting stability and instability properties - namely, black strings and branes. 
In fact, it was in the context of black strings that the first unstable vacuum black object was found, by Gregory and Laflamme \cite{Gregory:1993vy}, and it was this work that inspired much of the later work on instabilities of asymptotically flat spacetimes. Later, Gubser and Mitra \cite{Gubser:2000ec} conjectured that this instability could be understood in terms of the thermodynamics of black brane, and in particular that \emph{any black string/brane which is thermodynamically unstable is also dynamically unstable to long-wavelength perturbations}. This conjecture was recently proved by Hollands and Wald, using the canonical
energy method discussed below. Note that such a statement clearly fails if made about black holes instead of black branes, since even the Schwarzschild black hole is thermodynamically unstable.

Traditional approaches to studying the linear stability or instability of a particular solution involve writing out the linearised field equations (in some gauge) in that background, and then often invoking the symmetries of the solution to decompose an arbitrary perturbations into separate modes, and finally searching for a mode which grows exponentially in time. Although this approach has been successful in a number of cases (\cite{Dias:2011jg}, \cite{Dias:2009iu}, \cite{Dias:2010eu}, \cite{Dias:2010maa}), it would be nice to have a more general method, which could be applied to \emph{any} stationary background (regardless of symmetries). In addition, particularly when searching for an instability, it appears that the above method may be unnecessarily difficult, since it involves finding the form of an unstable mode \emph{exactly}. On the other hand, in order to prove the existence of an instability, we might hope that we can just look at perturbations which are sufficiently close to this unstable mode, since if we actually evolved these perturbations we would surely see the instability.

One approach which captures some of these desireable qualities was recently found in \cite{Figueras:2011he}, which is based on \emph{local penrose inequalities}. The idea is to consider the relationship between mass, angular momentum and horizon area for the family of solutions of interest. If we consider perturbations to the initial data for such a solution, we can calculate the corresponding perturbations to the mass, angular momentum and horizon area. If we evolve this perturbed data forwards in time then we know two things: first that the horizon area can only increase (by the second law), and second that the mass can only decrease (as gravitational radiation carries away energy). We now ask the question: given these restrictions, can the perturbed solution ever return to the original family of solutions? If there exists a perturbation for which the answer is ``no'', then we have demonstrated existence of an instability for this family of solutions. The inequality which results from the considerations above in the positive is called a ``local penrose inequality'', and violating it demonstrates instability.

Clearly this approach is superior to the one previously described, in that we do not need to find the exact form of the unstable perturbation. In particular, in order to apply this method in practice we need to construct a local penrose inequality for the family of solutions under consideration, then calculate the variations of angular momentum, mass and horizon area for some chosen perturbations, and look for a violation of the inequality. On the other hand, the first law constrains all such perturbations to leading order, and so we need to use second order perturbation theory in order to find a violation.\footnote{This may appear strange, because we would expect unstable behaviour to manifest itself in growing modes of the \emph{linearised} equations of motion. However, the particular combination of second order perturbed quantities which need to be considered will actually depend only0 on the first order perturbations.} In addition, this approach does not seem to be able to tell us anything about \emph{stability}, since even if we could show that all perturbations satisfy a local Penrose inequality, this is only a necessary (but not sufficient) condition for stability. 

A closely related approach was recently developed by Hollands and Wald in \cite{Hollands:2012sf}, which resolves some of these issues. The approach is based on the canonical energy, and provides us with a quadratic form, $\mathcal{E}$ (which depends on the background spacetime), evaluated on our initial data perturbations. It has the property that its value can only decrease in the future, together with certain other properties such as gauge-invariance, and after taking account of stationary perturbations, this allows us to use $\mathcal{E}$ to discuss stability and instability. In particular, if we can find a perturbation making $\mathcal{E}$ negative, then this perturbed spacetime can never return to the original spacetime, while if we can show that $\mathcal{E} \geq 0$ for all perturbations then we have established linear stability.

The advantages of this approach should be clear: as in the local Penrose inequality approach, we do not need to find the precise form of an unstable mode, but instead only need to evaluate $\mathcal{E}$ on various test perturbations. However, this time we can potentially use this approach to say something about stability as well as instability. We can also give a (lengthy) expression for $\mathcal{E}$ on a general background spacetime, in contrast to the slightly more ``case by case'' approach necessary in the local Penrose inequality approach.

The purpose of this paper is to demonstrate that the definition of $\mathcal{E}$ - henceforth called the \emph{canonical energy} (though it is really the second variation of the energy) - can be extended to cover gravity coupled to a wide variety of matter models, whilst retaining the properties which made it suitable for the study of stability and instability. Since the original definition was given in terms of a generic, diffeomorphism-invariant Lagrangian this may seem a straightforward task, however it is complicated by two factors. First, the canonical energy is intimately related to the various conserved charges of the theory, and since we include gauge fields in our Lagrangian we add additional charges. Second, the key property possessed by $\mathcal{E}$ is its decay, and in order to show this we need to demonstrate that the flux of canonical energy across null infinity is positive. We therefore require an appropriate definition of asymptotic flatness in the presence of our matter fields, which we give. Finally, we will also extend the proof of the Gubser-Mitra conjecture to the case of charged black strings in the presence of our matter fields.

To be precise, we will study the Lagrangian, on a $D$-dimensional manifold $\mathcal{M}$:
\begin{equation}
 \mathcal{L} = *R - \frac{1}{2}F_{BC}(\phi)\upd\phi^B \wedge * \upd\phi^C - *V(\phi) - \frac{1}{2}G_{IJ}(\phi)\upd A^I \wedge\upd A^J
\label{lagrangian}
\end{equation}
where the $\phi^A$ are scalar fields and the $A^I$ are $p$-form gauge fields. We require the matrices $F_{AB}$ and $G_{IJ}$ to be invertible and positive-definite, and also that $V(0) = 0$ is a minimum of the scalar field potential. This means that we can have decay of the scalar fields near infinity - if it were not the case, then we could just make a field redefinition corresponding to a constant shift in the scalar fields.
Some additional constraints on the coefficients $F_{BC}$ and $G_{IJ}$ will be needed in order to have a well-defined notion of asymptotic flatness, and these will be discussed below. This Lagrangian is clearly very general, and solutions to this theory will possess a wide range of charges associated with the gauge fields.

\section{The Spacetimes Under Consideration}

We will consider either asymptotically flat, or asymptotically Kaluza-Klein, stationary and axisymmetric solutions to the theory defined by the Lagrangian (\ref{lagrangian}). The appropriate definitions of ``asymptotically flat'' and ``asymptotically Kaluza Klein'' will be given later. We will consider solutions possessing a non-extremal event horizon with compact cross sections consisting of a single connected component, though we do not require further restrictions on the horizon topology. This includes black holes and (in the asymptotically Kaluza Klein case) black strings, as well as black rings, and potentially more complicated objects.
\\[10pt]
{\bf Stationary, Axisymmetric Spacetimes}
We should be careful in our definitions of ``stationary'' and ``axisymmetric'' solutions, since the Lagrangian (\ref{lagrangian}) admits a more general class of symmetries than isometries. In fact, the symmetries are generated by generalised Killing vector fields, consisting of a pair $(X, \{\Lambda^I\})$, where $X$ is a vector field and the $\{\Lambda^I\}$ are $(p-2)$-forms, satisfying
\begin{equation}
 \begin{split}
  \mathscr{L}_X g &= 0\\
  \mathscr{L}_X \phi^B &= 0\\
  \mathscr{L}_X A^I + \upd \Lambda^I &= 0
 \end{split} \end{equation} In other words, we should allow the gauge fields to change under the isometry generated by the Killing vector field $X$, but only by a gauge transformation.

We now define a stationary spacetime as follows: there exists an asymptotically timelike, generalised Killing field $(t, \{\Lambda^I_{(t)}\})$. Similarly, an axisymmetric spacetime possesses $(D-3)$ mutually commuting asymptotically spacelike generalised Killing vector fields $(\psi^i, \{\Lambda^I_{(\psi^i)}\})$, with the $\psi^i$ possessing closed orbits. Finally, if our spacetime is asymptotically Kaluza-Klein, then there will be additional spacelike generalised Killing vector fields $(X^j, \{\Lambda^I_{(X^j)}\})$ associated with the toroidal directions.

We also restrict consideration to solutions which possess non-extremal Killing horizons, so that there is a generalised Killing vector field $(K, \{\Lambda^I_{K}\})$, with
\begin{equation}
 K = t - \sum_i \Omega_i \psi^i - \sum_j k_j X^j
\label{killing}
\end{equation}
where $K$ is normal to the horizon and vanishes on the bifurcation surface. Here, the $\Omega_i$ and $k_j$ are constants, called the angular velocities and linear velocities (in the compact, Kaluza-Klein directions) of the horizon.
\\[10pt]
We wish to allow for magnetically charged solutions, and will therefore not demand that the gauge fields $A^I$ are globally defined. However, we will require the electrostatic potentials, $\imath_K A^I$ to be globally defined. If $A^I$ is a $p$-form, then the associated electrostatic potentials are $(p-1)$-forms, and so topological obstructions to their (global) existence are different from the obstructions which would allow the construction of magnetically charged solutions. Under suitable assumptions on the topology of the region of outer communication, we could therefore ensure that this condition is met. In any case, the electrostatic potential is globally defined in all the spacetimes discussed in this article. If the solution has electric charge but no magnetic charge, then the gauge fields are themselves globally defined and so the electrostatic potentials are trivially globally defined. On the other hand, if the solution is magnetically charged but not electrically charged, then the electrostatic potential will vanish.

Our asymptotic flatness conditions will include a certain fall-off rate for the gauge fields, $A^I$, which can be found in section \ref{asymptotics section}. These conditions will allow us to consider both electrically and magnetically charged black holes (in any dimension), as well as most of the \emph{dipole black rings} of \cite{Emparan:2004wy}, and their black brane counterparts. However, there are certain spacetimes which our conditions are too strict to contain. In fact, if the solution is magnetically charged, with field strength associated with a gauge field proportional to the volume form on some $(p+1)$-cycle, then our asymptotic conditions mean that we must have $p+2 \geq D/2$. While all magnetically charged black holes obey this restriction, a magnetically charged dipole black ring in seven dimensions, which is charged with respect to a one-form potential, does not. On the other hand, the dual electrically charged dipole black ring in $D = 7$ \emph{is} permissible\footnote{It is not true that, for all charged black objects, either the electrically charged solution or its magnetic dual satisfies our asymptotic conditions. Even if this were the case, if spacial cross sections of the horizon contain both nontrivial $(p+1)$-cycles and $(D-p-1)$-cycles, then there may exist solutions with both electric and magnetic charge corresponding to single gauge field, and so we lose generality by restricting to solutions with only an electric or a magnetic charge. Nevertheless, many charged configurations are allowed.}.

From this point onwards, we will assume that the gauge fields have been put into a gauge such that all of the $\{\Lambda^I_X\}$ mentioned above vanish, i.e. such that the gauge fields are in fact stationary and axisymmetric. This is possible in all the known solutions, though there are possible topological obstructions to the existence of such a gauge in the general case. Nevertheless, we make this assumption for simplicity, and keep in mind that the $\Lambda^I$ can be easily reintroduced.

\section{First Variations and the First Law}

In this section we develop the formalism necessary (see \cite{Crnkovic:1986ex}, \cite{Compere:2007vx}) to define the canonical energy, and also use this to find the form of the first law of black hole mechanics in our setting. To save on notation, we set $S = (g, \{\phi^B\}, \{ A^I\})$, so that a solution is given by specifying $S$. We first consider varying our Lagrangian (\ref{lagrangian}), finding
\begin{equation}
 \delta\mathcal{L} = E(\delta S; S) + \upd\theta(\delta S; S)
\label{lagrangian varied}
\end{equation}
where $E(\delta S; S)=0$ for all $\delta S$ provide the field equations, and $\theta$ is a boundary term. Explicitly, we find that
\begin{equation}
 \theta(\delta S; S) = \imath_v\epsilon - F_{BC}\delta\phi^B*\upd\phi^C - G_{IJ}\delta A^I\wedge*\upd A^J \label{theta}
\end{equation}
where $\epsilon$ is the volume form associated with the background metric $g$, and the vector $v$ is given by
\begin{equation}
 v^\mu = 2g^{\mu\rho}g^{\nu\sigma}(\nabla_\sigma\delta g_{\nu\rho} - \nabla_\rho\delta g_{\nu\sigma})
\end{equation}
The \emph{symplectic current} is a function of two variations, and is defined by
\begin{equation}
 \omega(\delta_1 S, \delta_2 S; S) = \delta_1\theta(\delta_2 S; S) - \delta_2\theta(\delta_1 S;S)
\end{equation}
Here we are to think of our fields $(g, \{\phi^B\}, \{A^I\})$ as functions of two parameters, say $\lambda$ and $\mu$, and define
\begin{equation}
 \begin{split}
  \delta_1 &\leftrightarrow \frac{\partial}{\partial\lambda}\Big|_{\lambda=\mu=0} \\
  \delta_2 &\leftrightarrow \frac{\partial}{\partial\mu}\Big|_{\lambda=\mu=0}
 \end{split}
\end{equation}
Since partial derivatives commute, the symplectic current defined above really is a function of the first order perturbations, and not, say, $\delta_1\delta_2 g$. We also find that
\begin{equation}
 0 = \delta_1\delta_2\mathcal{L} - \delta_2\delta_1\mathcal{L} = \upd\omega(\delta_1 S, \delta_2 S; S) + \delta_1 E(\delta_2 S; S) - \delta_2 E(\delta_1 S; S)
\end{equation}
When the linearised field equations are satisfied by both perturbations, i.e. $\delta_1 E = \delta_2 E = 0$, the symplectic current is closed, that is, $\upd\omega(\delta_1 S, \delta_2 S; S) = 0$. The \emph{symplectic form} is the integral of the symplectic current over a $(D-1)$-dimensional surface $\Sigma$:
\begin{equation}
 W_\Sigma(\delta_1 S, \delta_2 S;S) = \int_\Sigma\omega(\delta_1 S, \delta_2 S;S)
\end{equation}
If the linearised field equations are satisfied, then the symplectic form is conserved between homologous surfaces, ince $\omega$ is closed.

Since diffeomorphisms and gauge transformations are generated by a pair $(X, \{\Lambda^I\})$, where $X$ is a vector field, and the $\Lambda^I$ are a set of ($p-1$)-forms, we define the \emph{Noether current} associated to such a pair by:
\begin{equation}
 \begin{split}
 \mathcal{J}_{(X, \Lambda)} &= \theta(\mathscr{L}_X g, \{\mathscr{L}_X \phi^B\}, \{\mathscr{L}_X A^I + \upd \Lambda^I\}; S) - \imath_X \mathcal{L}\\
 &= \theta_{\text{GR}}(\mathcal{L}_X g; g) - \imath_X\mathcal{L}_{\text{GR}}(g) - \frac{1}{2}F_{BC}\mathscr{L}_X\phi^B*\upd\phi^C - \frac{1}{2}F_{BC}\upd\phi^B\wedge\imath_X*\upd\phi^C + \imath_X*V \\
 & \; - \frac{1}{2}G_{IJ}\imath_X\upd A^I\wedge*\upd A^J - (-1)^p\frac{1}{2}G_{IJ}\upd A^I\wedge\imath_X*\upd A^J + (-1)^{p-1}\left(\Lambda^I + \imath_XA^I\right)\wedge\upd(G_{IJ}*\upd A^J) \\
 & \; - \upd\left(\left(\Lambda^I + \imath_XA^I\right)G_{IJ}*\upd A^J\right) \label{noether} \end{split} \end{equation}
where $\mathcal{L}_{\text{GR}}$ is the Einstein-Hilbert Lagrangian, and $\theta_{\text{GR}}$ is related to it by (\ref{lagrangian varied}). It can be shown \cite{Iyer:1995kg} that, for a general diffeomorphism-invariant theory,
\begin{equation}
 \mathcal{J}_{(X, \Lambda)} = C_{(X, \Lambda)} + \upd Q_{(X, \Lambda)}
\label{current}
\end{equation}
where $C_{(X, \Lambda)}$ vanishes whenever the constraints of the theory hold, and $Q_{(X, \Lambda)}$ is the Noether charge associated with $(X, \{\Lambda^I\})$. In our case, we can see from (\ref{noether}) that \begin{equation}
 (Q_{(X, \Lambda)})_{\mu_1 \ldots \mu_{D-2}} = - \frac{1}{2}\nabla^\nu X^\rho \epsilon_{\nu\rho\mu_1\ldots \mu_{D-2}} -\left(G_{IJ}\left(\Lambda^I + \imath_XA^I\right)\wedge*\upd A^J\right)_{\mu_1\ldots\mu_{D-2}}
\label{noethercharge}
\end{equation}
where all the matter terms in (\ref{noether}) except for the last one arise from the Hamiltonian, momentum and Gauss law constraints when pulled back to a spacelike hypersurface\footnote{The easiest way to see this is to take $X = \partial/\partial t$ and assume that it is hypersurface-orthogonal, and then to pull back equation (\ref{noether}) to the hypersurface $t = \text{constant}$. Then we recognise these terms as the Hamiltonian and Gauss law constraints. In order to see the momentum constraint, we need to decompose $X$ into a lapse and shift in the usual way.}.

On the other hand, a straightforward calculation shows that variations of the Noether current satisfy \begin{equation}
 \delta\mathcal{J}_{(X, \Lambda)} = -\imath_XE(\delta S) + \omega(\delta S, \mathscr{L}_XS + \delta_\Lambda S) + \upd\imath_X\theta(\delta S) + \delta_\Lambda\theta(\delta S)
\label{deltaj}
\end{equation}
where the perturbation $\delta_\Lambda$ is defined by
\begin{equation}
 \delta_\Lambda S = (0, 0, \{\Lambda^I\})
\end{equation}
Perturbing (\ref{theta}) in this way, we see that in fact $\delta_\Lambda\theta(\delta S)$ = 0. Combining (\ref{current}) and (\ref{deltaj}), and for perturbations satisfying the linearised field equations we obtain 
\begin{equation}
 \omega(\delta S, \mathscr{L}_X S + \delta_\Lambda S; S) = \upd\left(\delta Q_{(X, \Lambda)} - \imath_X\theta(\delta S)\right)
\label{omegaexact}
\end{equation}


\subsection{The First Law}
\label{first law subsection}
In order to obtain the first law, we simply evaluate the symplectic current $W_{\Sigma}(\delta S, \mathscr{L}_K S)$, where $K$ is the horizon Killing field described above, and $\Sigma$ is a Cauchy surface for the exterior region, with boundaries on the bifurcation surface and at spatial infinity. From (\ref{omegaexact}) it appears that we will only pick up contributions from these boundaries. However, we want to include magnetically charged objects in our discussion; the easiest way to do this is to work in multiple patches, and to allow the gauge fields $A^I$ to change (by a gauge transformation) along the edges of each patch. We will then pick up contributions from each of these edges, which we denote collectively by $E$. Thus we find
\begin{equation}
 \begin{split}
 W_{\Sigma}(\delta S, \mathscr{L}_K S + \delta_\Lambda S; S) &= \int_\infty\left(\delta Q_{(K, \Lambda)} - \imath_K\theta(\delta S; S)\right) + \int_B\left(\delta Q_{(K, \Lambda)} - \imath_K\theta(\delta S;S)\right) \\
 & \phantom{A} + (-1)^p\int_{E\setminus\partial\Sigma} G_{IJ}\Delta\delta A^I\wedge\imath_K*\upd A^J \end{split}
\label{firstlaw1}
\end{equation}
where $\Delta A^I$ is the change in $A^I$ across the edge $E$. Properly, we should ensure that the patches on which the $A^I$ are smooth overlap in some open region, such that, in the overlap of any two patches, the
difference between the two gauge fields is a gauge transformation. However, for our purposes we will only need to consider the change in the gauge fields across some edge between the patches (which we imagine as lying within an open region as described above).
\\[8pt]
{\bf Einstein-Hilbert Terms}
We begin by evaluating the contributions to (\ref{firstlaw1}) arising from the Einstein-Hilbert terms. Using (\ref{killing}), as in \cite{Hollands:2012sf} we identify the terms
\begin{equation}
\begin{split}
 \int_\infty \left(\delta Q_{\text{GR}(t)} - \imath_t\theta_{\text{GR}}(\delta g;g)\right) &= \delta M_{\text{GR}} \\
 \int_\infty \left(\delta Q_{\text{GR}(\Omega_i\psi^i)} - \Omega_i\imath_{\psi^i}\theta_{\text{GR}}(\delta g;g)\right) &= \Omega_i\delta J^i_{\text{GR}} \\
  \int_\infty \left(\delta Q_{\text{GR}(k_j X^j)} - k_j\imath_{X^j}\theta_{\text{GR}}(\delta g; g)\right) &= k_j\delta P^j_{\text{GR}} \\
 \int_B \left(\delta Q_{\text{GR}(K)} - \imath_K\theta_{\text{GR}; g}(\delta g; g)\right) &= 4\kappa\delta \mathcal{A}
\end{split}
\end{equation}
where the quantities $M_{\text{GR}}$, $J^i_{\text{GR}}$ and $\mathcal{A}$ are the usual ADM expressions for the mass, angular momenta and horizon area of an asymptotically flat spacetime, $P^j$ is the total momentum in the (compact) direction of the asymptotic Killing field $X^j$, and $\kappa$ is the surface gravity of the background spacetime. Note in particular that the second term in the integral for the horizon area vanishes, since $K = 0$ on $B$.
\\[8pt]
{\bf Scalar Field Terms}
Next we examine the contribution from the scalar fields. The terms arising from the scalar fields can be written as
\begin{equation}
\begin{split}
 \int_{\infty}F_{AB}\delta\phi^A\imath_t*\upd\phi^B &= \delta M_\phi \\
 \int_{\infty}\Omega_iF_{AB}\delta\phi^A\imath_{\psi^i}*\upd\phi^B &= \Omega_i\delta J^i_\phi \\
 \int_\infty k_jF_{AB}\delta\phi^A\imath_{X^j}*\upd\phi^B &= k_j\delta P^j_{\text{GR}} \\
\end{split}
\label{scalarcontribution}
\end{equation}
We recognise these expressions as the extra contributions to the mass, angular momenta and linear momenta arising from the scalar field. With the rate of fall-off we will impose on the scalar field, all of these terms vanish.
\\[8pt]
{\bf Gauge Field Terms: Electric Charges}
Finally, we examine the contributions from the gauge field. We find that
\begin{equation}
 \delta Q_{\text{gauge}(K, 0)} - \imath_K\theta_{\text{gauge}}(\delta S;S) = (-1)^pG_{IJ}\delta A^I\wedge\imath_K*\upd A^J - \imath_KA^I\wedge\delta\left(G_{IJ}*\upd A^J\right)
\label{gauge field contribution}
\end{equation}
The first term is analogous to the term examined above for scalar fields, and so may give contributions from the gauge field to the mass, angular momenta and linear momenta, although with our asymptotic flatness conditions all these contributions will vanish. The second term gives rise to electric charge, as we will show below. Note that \emph{physical} quantities, such as $\upd A$ should be smooth at the horizon, and therefore satisfy
\begin{equation}
 \imath_K \upd A^I \big|_B = 0
\end{equation}
but no such restriction must necessarily be made on \emph{unphysical} quantities. In particular, $\imath_K A^I$ need not vanish on the horizon. Now, using the fact that $(K, 0)$ is a generalised Killing vector field with $K = 0$ on $B$, we see that the quantities $\imath_K A^I$ are closed on $B$, since $\upd \imath_K A^I = \mathscr{L}_K A^I = 0$.

We need to calculate the integral
\begin{equation}
 \int_{\partial\Sigma}\imath_K A^I\wedge\delta\left(G_{IJ}*\upd A^J\right)
\label{electric1}
\end{equation}
Since the $\imath_K A^I$ are closed on $B$ we can write
\begin{equation}
 \imath_K A^I = \upd \lambda^I + h^I
\end{equation}
on $B$, where $h^I$ is a harmonic $(p-1)$ form with respect to the induced Laplace-Beltrami operator on $B$. The exact form $\upd\lambda^I$ does not contribute to the integral (\ref{electric1}) because of the linearised field equations for the gauge field. On the other hand, we can write $h^I = \Psi^{(I, \alpha)}\rho_{\alpha}$, where the $\Psi^{(I, \alpha)}$ are constants (they are the \emph{electrostatic potential differences}) and the $\rho_{\alpha}$ are dual to a set of non-homologous, non-contractible $(D-p-1)$ cycles of $B$, $T_\alpha$:
\begin{equation}
 \int_B \rho_\alpha\wedge \mu = \int_{T_\alpha} \mu
\end{equation}
for any $(D-p-1)$ form $\mu$. Thus, we can define the \emph{electric charges}
\begin{equation}
 \mathcal{Q}_{(I, \alpha)} = \int_{T_\alpha}G_{IJ}*\upd A^J
\end{equation}
We then see that the term (\ref{electric1}) is given by
\begin{equation}
 \Psi^{(I, \alpha)}\delta \mathcal{Q}_{(I, \alpha)}
\end{equation}

We now make some remarks on the above formulae. Note that we found the electric charges as integrals over the bifurcation surface, rather than as integrals over infinity. However, in some situations we are able to choose a different gauge, so that (\ref{electric1}) reduces to an integral over infinity alone; this is what is usually done in the Hamiltonian approach. Indeed, when this is possible we must get the same answer, since we can write (\ref{electric1}) as
\begin{equation}
 \int_\Sigma\upd\imath_KA^I\wedge\delta\left(G_{IJ}*\upd A^J\right)
\end{equation}
and this is clearly invariant under gauge transformations which preserve the stationarity and axisymmetry of the gauge fields. It is in this sense that the constants $\Psi^{(I, \alpha)}$ are potential differences: regarding $(\imath_KA^I + \Lambda^I)$ as a potential associated with the cycle labelled by $\alpha$, if this cycle can be deformed to infinity, then $\Psi^{(I, \alpha)}$ is precisely the difference in this potential between the horizon and infinity. On the other hand, such deformations do not always exist, since the horizon may possess non-contractible $(D-p-1)$-cycles which do not exist on the surface at infinity. For example, spatial cross sections of the horizon of a black ring in five dimensions are homeomorphic to $S^2\times S^1$. They possess non-contractible 2- and 1-cycles, but the surface at infinity is just a 3-sphere. In other words, there are cycles which cannot be continuously deformed to infinity, since they pass through the middle of the black ring. Note also, from this example, that a black ring in five dimensions may possess electric charge with respect to a 2-form gauge field\footnote{This is an example of a \emph{dipole charge}, for which the corresponding gauge field decays like a dipole near infinity. See \cite{Copsey:2005se} for a more detailed discussion and further examples.}, since then $*\upd A$ is a 2-form, whereas a black hole in five dimensions (with horizon topology $S^3$) can possess electric charge with respect to a 1-form gauge field, as in this case $*\upd A$ is a 3-form.

Temporarily reinstating the $\Lambda^I$ (i.e. replacing $\imath_K A^I$ with $(\imath_K A^I + \Lambda^I)$ in the above formulae), we note that, in order for the electrostatic potential differences to be nonzero in a gauge where the $A^I$ decay appropriately at infinity, we must have $(\imath_KA^I + \Lambda^I)\Big|_B \neq 0$. However, $K = 0$ on the bifurcation surface, $B$. Thus we see that, if the $\Lambda^I$ vanish on $B$, then the gauge field must diverge on $B$. This should not come as a complete surprise: for the Reissner-N\"ordstrom black hole we can pick a gauge in which
\begin{equation}
 A = \frac{r_+}{r}\upd t
\end{equation}
and this diverges on the bifurcation surface, since the norm of $\upd t$ diverges there. This differs from the usual gauge choice:
\begin{equation}
 A = \left(1-\frac{r_+}{r}\right)\upd t
\end{equation}
which is smooth at the horizon, but does not decay near infinity. Of course, both potentials describe the same physical system, and it is easy to check that $\upd A$ is smooth on the horizon and decays near infinity in both cases. On the other hand, it may be possible to choose the $\Lambda^I$ appropriately so that the $A^I$ do not diverge at the horizon, even if they are chosen to vanish near infinity, but this issue will not be discussed here.
\\[8pt]
{\bf Gauge Field Terms: Magnetic Charges}
Finally, we need to deal with the term
\begin{equation}
 (-1)^p\int_{E\setminus\partial\Sigma} G_{IJ}\Delta\delta A^I\wedge\imath_K*\upd A^J
\end{equation}
which is the only term arising from integrals over the edges of the patches, since the $A^I$ change discontinuously between the patches, but the $\upd A^I$ are smooth and so (by assumption) are the $\imath_K A^I$. We know that
\begin{equation}
 \upd\Delta\delta A^I = \upd\left(G_{IJ}\imath_K*\upd A^I\right) = 0
\end{equation}
where the first equality holds because $\Delta\delta A^I$ is a gauge transformation and the second holds due to the field equations and because $K$ is a generalised Killing vector. Moreover, we can assume that $\Delta\delta A^I$ is closed but not exact, since any exact part can be absorbed into the gauge fields. We cannot use the usual Poincar\'e duality as we were able to in the case of electric charge, as the manifold $E$ has a boundary (although if we include a boundary at infinity, then $E$ is compact). Instead we must appeal to Lefschetz duality, which provides us with a duality between elements of the $p$-th cohomology group of $E$, and elements of the $(D-p-2)$-th relative homology group of $(E, \partial E)$. In particular, we can write
\begin{equation}
 \Delta\delta A^I = (-1)^p\sum \delta \mathcal{P}^{(I, \beta)} \sigma_\beta
\label{edge gauge field perturbation}
\end{equation}
where the $\mathcal{P}^{(I, \beta)}$ are constants, and the $\sigma_\beta$ satisfy
\begin{equation}
 \int_E \sigma_\beta\wedge \mu = \int_{C_\beta}\mu
\end{equation}
where $C_\beta$ is a $(D-p-2)$-chain with boundaries on $\partial E$, and $\mu$ is any $(D-p-2)$-form. Moreover, if $\mu$ is closed and vanishes on $\partial E$ then the above equation is independent of our choice of chain $C_\beta$, within the chosen relative homology class. So, if we identify the magnetostatic potential differences as
\begin{equation}
 \Phi_{(I, \beta)} = \int_{C_\beta}G_{IJ}\imath_K*\upd A^I
\label{magnetostatic}
\end{equation}
then, since the integrand is closed, vanishes on $B$ and tends to zero near infinity, this actually only depends on the the relative homology class of $C_\beta$. Then we have
\begin{equation}
 (-1)^p\int_{E\setminus\partial\Sigma} G_{IJ}\Delta\delta A^I\wedge\imath_K*\upd A^J = \Phi_{(I, \beta)}\delta \mathcal{P}^{(I, \beta)}
\end{equation}

We can now show that the charges $P^{(I, \beta)}$ are actually magnetic charges. Since $\Delta\delta A$ is a member of the $p$th cohomology class of $E$, and we assume that all charges arise from the presence of the horizon, there must be some non-contractable (in $E$) $p$-cycle, $\tilde{T}_\beta$ of $B\cap E$ corresponding to $\sigma_\beta$, i.e.
\begin{equation}
 \int_{\tilde{T}_\beta} \sigma_\gamma = \delta^\beta_\gamma
\end{equation}
If we assume that these cycles arise from the intersection of nontrivial $(p+1)$-cycles of $B$ with $E$, which we label $\tilde{T}^*_{\beta}$, then we easily see that
\begin{equation}
 \delta \mathcal{P}^{(I, \beta)} = \int_{\tilde{T}_\beta}\Delta\delta A^I = \int_{\tilde{T}^*_\beta}\upd \delta A^I
\end{equation}
and so the magnetic charges $P^{(I, \beta)}$ are just what we expect them to be: integrals of the $\upd A^I$ over the non-contractable (in $\Sigma$) $(p+1)$-cycles of the horizon, that is
\begin{equation}
 \mathcal{P}^{(I, \beta)} = \int_{\tilde{T}^*_\beta}\upd A^I
\end{equation}

Let us see how this works in the familiar case of a magnetically charged Reissner-N\"ordstrom black hole in four dimensions, with magnetic charge $\mathcal{P}$. The metric and field strength are
\begin{equation}
\begin{split}
 \upd s^2 &= -f(r)\upd t^2 + f^{-1}\upd r^2 + r^2(\upd\theta^2 + \sin^2\theta\upd\phi^2)\\
 \upd A &= \frac{1}{4\pi}\mathcal{P}\sin\theta\upd \theta\wedge\upd\phi
\end{split}
\end{equation}
where $f(r) \rightarrow 1$ at infinity and $f(r) \rightarrow 0$ at $r=r_+$, which is the position of the horizon, and $K = \partial/\partial t$. If we take $\Sigma$ to be a constant time slice (with boundary at $r=r_+$), we can take
\begin{equation}
 A = \left\{
  \begin{array}{ll}
    \frac{1}{4\pi}\mathcal{P}(1-\cos\theta)\upd\phi & \quad 0 \leq \theta \leq \pi/2\\
    \frac{1}{4\pi}\mathcal{P}(-1-\cos\theta)\upd\phi & \quad \pi/2 < \theta \leq \pi
  \end{array} \right.
\end{equation}
then $E$ is the equatorial plane in $\Sigma$, which is homeomorphic to the plane with a disc removed. Note that $\Delta A = \frac{1}{2\pi}\mathcal{P}\upd\phi$, which, as promised, is a member of the first cohomology group of $E$. An example of a chain in the dual relative homology class is a line of constant $\phi$, stretching from $r=r_+$ to infinity. The Hodge dual to the field strength is
\begin{equation}
 *\upd A = \frac{\mathcal{P}}{4\pi}\frac{1}{r^2}\upd t\wedge \upd r
\end{equation}
and we see that the magnetostatic potential difference is
\begin{equation}
 \int_{r_+}^\infty \frac{\mathcal{P}}{4\pi}\frac{1}{r^2} \upd r = \frac{\mathcal{P}}{4\pi r_+}
\end{equation}
which is the result expected from performing a duality transformation on the corresponding electrically
charged black hole.
\\[10pt]
Returning to (\ref{firstlaw1}), and noting that the left hand side vanishes since $(K, \{\Lambda\})$ is Killing, we can now evaluate all the terms to find the \emph{first law} in this setting:
\begin{equation}
 0 = W_{\Sigma}(\delta S, \mathscr{L}_K S + \delta_\Lambda S) = \delta M - 4\kappa\delta\mathcal{A} - \Omega_i\delta J^i - k_j\delta P^j - \Psi^{(I, \alpha)}\delta \mathcal{Q}_{(I, \alpha)} - \Phi_{(I, \beta)}\delta \mathcal{P}^{(I, \beta)}
\label{firstlaw}
\end{equation}
where all indices are summed over, and the combined total mass (including possible contributions from the matter fields), linear momenta and angular momenta are understood.

\section{Second Variations and the Canonical Energy}
We now take a second variation of (\ref{firstlaw}). Note that $W_{\Sigma}(\delta S, \mathscr{L}_K S + \delta_\Lambda S; S)$ depends on the background fields $S$ as well as their first variations. However, since $(K, \{\Lambda\})$ is Killing in the background, we obtain
\begin{equation}
 W_{\Sigma}(\delta S, \mathscr{L}_K \delta S; S) = \delta^2 M - 4\kappa\delta^2\mathcal{A} - \Omega_i\delta^2 J^i - k_j\delta^2 P^j - \Psi^{(I, \alpha)}\delta^2 \mathcal{Q}_{(I, \alpha)} - \Phi_{(I, \beta)}\delta^2 \mathcal{P}^{(I, \beta)}
\label{secondvariation}
\end{equation}
Note that the surface gravity, angular velocities, linear velocities in the compact directions, and electrostatic and magnetostatic potential differences are all defined in the background spacetime and so are not affected by this variation. In particular, $\delta_\Lambda S$ is actually independent of $S$, and so vanishes when we take another variation. In addition, we notice that the right hand side of (\ref{secondvariation}) appears to depend on the second order perturbations to our fields $S$, and indeed, each individual term will depend on both first and second order perturbations. However, all terms depending on second order perturbations must cancel in the sum, as the left hand side depends only on first order perturbations. In practice, this gives us an easy way to evaluate $W_{\Sigma}(\delta S, \mathscr{L}_K \delta S; S)$ by evaluating second order perturbations.

We now use (\ref{killing}) once again, and write
\begin{equation}
 W_{\Sigma}(\delta S, \mathscr{L}_K \delta S; S) = \mathcal{E}(\delta S; S) + \Omega_iW_{\Sigma}(\delta S, \mathscr{L}_{\psi^i} \delta S; S) + k_iW_{\Sigma}(\delta S, \mathscr{L}_{k^i} \delta S; S)
\label{canonicaldefinition1}
\end{equation}
where we have finally defined the \emph{canonical energy}:
\begin{equation}
 \mathcal{E}(\delta S; S) = W_\Sigma(\delta S, \mathscr{L}_t \delta S; S)
\label{canonicaldefinition2}
\end{equation}
The second two terms on the right hand side of (\ref{canonicaldefinition1}) will cause us problems when we discuss the positivity properties of $\mathcal{E}$, associated with the phenomena of superradiance. To avoid these, we assume from now on that our perturbations are axisymmetric (at least in the planes associated with nonzero angular velocities) and invariant under the action of the Killing fields on the compact manifold (at least in the directions associated with nonzero linear momenta).

\section{Gauge Choice}
Since the canonical energy is defined as an integral over some Cauchy surface for the exterior region of a black hole (or ring, or membrane), it seems reasonable to demand that the interior boundary of this surface coincides with position of the bifurcation surface in the perturbed spacetime as well as in the background one. We should be careful, however: suppose we have a family of spacetimes
\begin{equation}
 g_{\mu\nu} = g_{\mu\nu}(\lambda)
\end{equation}
where $g_{\mu\nu}(0)$ is the stationary, black hole spacetime whose stability we are trying to determine, and $\lambda > 0$ corresponds to a perturbed spacetime. Then there is no guarantee that a horizon exists for $g_{\mu\nu}(\lambda)$ for any $\lambda \neq 0$, and, even if one does, since these spacetimes are not in general stationary, the position of this horizon will be difficult to determine.

However, in \cite{Hollands:2012sf} it was shown that, when considering linearised perturbations, there is always a null surface in the perturbed spacetime with compact spatial cross sections, whose expansion vanishes to first order in $\lambda$. Since we will prove below that the perturbed expansion is constant along the generators of this surface, it must coincide with the position of the horizon - whenever it exists - to first order in $\lambda$.

In order to prove this, we first work in \emph{Gaussian normal co-ordinates}, i.e. in a neighbourhood of the horizon we set
\begin{equation}
 \upd s^2(\lambda) = 2\upd u \upd r - r^2 \alpha(\lambda)\upd u^2 - r \upd u \beta(\lambda) + \mu_{\alpha\beta}(\lambda)\gamma^\alpha(\lambda)\gamma^\beta(\lambda)
\label{gaussian}
\end{equation}
where the one-forms $\beta$ and $\gamma^\alpha$ (whose dependence on the other coordinates is suppressed) are orthogonal to the normal bundle of the joint level sets
of $u$ and $r$, i.e.
\begin{equation}
\begin{split}
 0 &= \beta(\lambda)\left(\frac{\partial}{\partial u}\right) = \beta(\lambda)\left(\frac{\partial}{\partial r}\right) \\
 0 &= \gamma^\alpha(\lambda)\left(\frac{\partial}{\partial u}\right) = \gamma^\alpha(\lambda)\left(\frac{\partial}{\partial r}\right)
\end{split}
\end{equation}
The surface $B = \{r=u=0\}$ is the bifurcation surface of a black hole, when $\lambda = 0$, and we define the vector field $n = (\partial/\partial u)$ Next we consider the one parameter family of diffeomorphisms $\phi_s$ generated by the vector field
\begin{equation}
 X = \frac{1}{\sqrt 2}f \frac{\partial}{\partial r}
\end{equation}
where $f$ is some smooth function. It can be shown \cite{Andersson:2005gq} that the expansion, $\vartheta(s)$ of the surface $B(s) = \phi_s(B)$ in the outward going, future directed null vector $k$ (normalised by $g(k, \partial/\partial r) = 1$) satisfies
\begin{equation}
\begin{split}
 \frac{\upd}{\upd s}\vartheta(s)\big|_{s = 0} &= -D^a D_a f + \beta^a D_a f + \frac{1}{2}\left(R(\mu) - \frac{1}{2}\beta^a\beta_a + D^a\beta_a - \frac{1}{2}(\mathscr{L}_{(\partial/\partial u)}\mu_{ab})(\mathscr{L}_{(\partial/\partial u)}\mu^{ab}) \right.\\
 & \left.- 2T_{ab}\left(\frac{\partial}{\partial u}^a\right)k^b\right)f \\
 &= C(f)
\end{split}
\end{equation}
where $D_a$ is the covariant derivative associated with $\mu_{ab}$. Since the null energy condition holds, we can repeat the argument of \cite{Hollands:2012sf} to conclude that the operator $C$ has strictly positive principle eigenvalue, and so we can uniquely solve the equation
\begin{equation}
 C(f) = -\delta\vartheta|_B
\end{equation}
and so, by means of a gauge transformation, we can set $\delta\vartheta|_B = 0$. In addition, we can use the remaining gauge freedom to set $\delta \epsilon_B = 0$, where $\epsilon_B$ is the induced volume form on the surface $B$, at least whenever the perturbed horizon area vanishes.


There is also gauge freedom in the variations of the gauge fields $\delta A^I$. We use this to impose the following condition: The pull-back to the future horizon of $\imath_n \delta A^I$ is closed, where $n$ is the future-directed null geodesic generator of the horizon. As in the case of gravitational perturbations, the imposition of this condition leaves a large amount of gauge-freedom.


\section{Properties of the Canonical Energy}
We wish to establish the following properties of $\mathcal{E}$, which are the key properties we use to connect the positivity properties of $\mathcal{E}$ to stability and instability:
\begin{enumerate}
 \item $\mathcal{E}(\delta S; S)$ is conserved in the sense that its value is the same on every surface $\Sigma$ with boundaries on spatial infinity and the bifurcation surface
 \item $\mathcal{E}(\delta S; S)$, for a given background $S$, can be viewed as a symmetric bilinear form on the initial value perturbations
 \item $\mathcal{E}(\delta S; S)$ is gauge-invariant (with respect to a certain class of gauge transformations)
 \item When restricted to a certain class of perturbations, $\mathcal{E} = 0$ if and only if the perturbation is towards a stationary solution
 \item For a certain class of perturbations, the fluxes of $\mathcal{E}(\delta S)$ across null infinity and across the horizon are non-negative
\end{enumerate}
We will find that properties 1 and 2 are easy to verify. We then need establish a technical lemma; properties 3 and 4 above will then be proved by fairly straightforward applications of this lemma, while property 5 requires a lot more work. Indeed, in order to establish the positivity of the flux of $\mathcal{E}$ through null infinity we first have to prove the linear stability of asymptotic flatness, and this allows us to read off the decay of the perturbations near null infinity (see section \ref{asymptotics section}).

The first two properties are established as follows: since $\omega(\delta_1 S, \delta_2 S; S)$ depends on the background gauge fields only through the terms $\upd A^I$, we have
\begin{equation}
 \mathscr{L}_t \omega(\delta_1 S, \delta_2 S; S) = \omega(\delta S_1, \mathscr{L}_t\delta S_2; S) - \omega(\delta S_2, \mathscr{L}_t\delta S_1; S) = \upd\imath_t \omega(\delta_1 S, \delta_2 S; S)
\end{equation}
so we see that
\begin{equation}
\mathscr{L}_t W_\Sigma(\delta S_1, \delta S_2; S) = W_\Sigma(\delta S_1, \mathscr{L}_t\delta S_2; S) - W_\Sigma(\delta S_2, \mathscr{L}_t\delta S_1; S) = \int_{\partial \Sigma} \imath_t \omega(\delta S_1, \delta S_2; S)
\end{equation}
The right hand side of this vanishes due to our asymptotic fall-off conditions near spacial infinity, and since $t$ is tangent to the bifurcation surface. Properties 1 and 2 then follow from setting $\delta S_1 = \delta S_2$. We next prove the technical lemma mentioned above:
\\[10pt] %
{\bf Lemma 1}: Let $\delta S$ solve the linearised field equations and our gauge conditions, and let $(\xi, \{\Theta^I\})$ be a pair consisting of a smooth vector field $\xi$, and a set of smooth, axisymmetric $(p-1)$-forms $\{\Theta^I\}$ such that
\begin{enumerate}[(i)]
 \item $\xi$ is tangent to the generators of the horizon at $B$
 \item The pull back of the $\Theta^I$ to $B$ are closed
 \item $(\xi, \{\Theta^I\})$ approaches an \emph{asymptotic symmetry} at spatial infinity, where an asymptotic symmetry is defined as a pair such that the contribution from the integral over infinity in (\ref{firstlaw1}) vanishes when we replace $(K, \Lambda)$ with $(Y, \Theta)$
 \item The first order perturbations (due to $\delta S$) to all of the charges arising in the first law, (\ref{firstlaw}), vanish 
\end{enumerate}
Then $W_\Sigma(\delta S, \mathscr{L}_\xi S + \delta_\Theta S) = 0$.
\\[10pt]%
A proof of this statement is given in \cite{Hollands:2012sf} in the case of metric perturbations. To extend this argument to our case we only need to show that the extra terms arising from the gauge and scalar fields in (\ref{firstlaw1}) at $B$ and along the edges $E$ vanish under the hypotheses above, as the contributions from the integrals over infinity vanish by assumption.

We first examine the terms in (\ref{firstlaw1}) (see also (\ref{gauge field contribution}) and (\ref{scalarcontribution}) for an idea of how these terms arise):
\begin{equation}
 \int_B\left(F_{BC}\delta\phi^B\imath_\xi*\upd\phi^C + (-1)^p G_{IJ}\delta A^I\wedge\imath_\xi *\upd A^J\right) \label{massterms}
\end{equation}
Let $\varphi: B \rightarrow \mathcal{M}$ be the embedding of $B$ in the manifold $\mathcal{M}$, and let $n$ be the null generator of the horizon, defined above (so $n = \partial/\partial u$ in Gaussian normal co-ordinates). Then we can calculate pull-backs to $B$:
\begin{equation}
\begin{split}
 \varphi_*\imath_n(*\upd\phi^B) &= *\varphi_* \mathscr{L}_n \phi^B \\
 \varphi_*\imath_n(*\upd A^I) &= *\varphi_* \left(\imath_n\upd A^I\right)
\end{split}
\end{equation}
where the Hodge star on the right hand side is the induced Hodge star on $B$, whereas on the left hand side it is the Hodge star associated with $\mathcal{M}$. However, since $n$ is a Killing field along the horizon, $\mathscr{L}_n\phi^B = 0$. In addition, since there is no energy flux along the horizon in the background, stationary solution, we have
\begin{equation}
 T_{\mu\nu}n^\mu n^\nu = 0
\label{horizon energy flux}
\end{equation}
which implies that $G_{IJ} \imath_n \upd A^I \cdot \imath_n\upd A^J = 0$ on the horizon, and so - since $G_{IJ}$ is non-degenerate - the $p$-forms $\imath_n \upd A^I$ are null at the horizon. Together with (\ref{gaussian}), this means that on the horizon, $\imath_n\upd A^I = \upd r \wedge \mu^I$ for some $(p-1)$-forms $\mu^I$. In particular, this means that the pull-back to $B$ of $\imath_n\upd A^I$ vanishes. Hence we find that both terms in (\ref{massterms}) vanish.

Next, we examine the term
\begin{equation}
 \int_B \left(\imath_\xi A^I + \Theta^I\right)\wedge\delta\left(G_{IJ}*\upd A^J\right) \label{eqn1} \end{equation}
We have $\xi = f n$, where $f$ is constant over the horizon. It can be shown (see \cite{Hollands:2012sf}) that, on the future horizon, $n = K/u$, and so if $\imath_K A^I$ is non-vanishing on $B$ the first term in (\ref{eqn1}) appears to diverge. If we define $B(\epsilon)$ as the surface $u = \epsilon$, $r=0$, then we have
\begin{equation}
\int_{B(\epsilon)} \left(f\imath_n A^I + \Theta^I\right)\wedge\delta\left(G_{IJ}*\upd A^J\right) = \frac{1}{\epsilon}\int_{B(\epsilon)} \left(f\imath_K A^I + \epsilon\Theta^I\right)\wedge\delta\left(G_{IJ}*\upd A^J\right)
\label{eqn2}
\end{equation}
We aim to show that the pull back to $B(\epsilon)$ of $(1/\epsilon)\left(f\imath_K A^I +
\epsilon\Theta^I\right)$ is closed in the limit $\epsilon \rightarrow 0$. This is true for the second term
by our conditions on the $\Theta^I$. For the first term, using the fact that $\epsilon$ is constant on $B(\epsilon)$ and that the gauge field is stationary, we have that
\begin{equation}
 \frac{1}{\epsilon}\upd\left(\imath_K A^I\right) = \frac{1}{\epsilon}\left(\mathscr{L}_K A^I - \imath_K \upd A^I\right) = - \imath_n \upd A^I
\end{equation}
This vanishes on $B$, as we have seen above. Thus, using the linearised field equations for the gauge fields, we can write
\begin{equation}
 \int_{B(\epsilon)} \left(f\imath_n A^I + \Theta^I\right)\wedge\delta\left(G_{IJ}*\upd A^J\right) = \int_{B(\epsilon)} \tilde{h}^I\wedge\delta\left(G_{IJ}*\upd A^J\right) + \mathcal{O}(\epsilon) \end{equation}
where $\tilde{h}^I$ is a harmonic form on $B(\epsilon)$. We can decompose this as before, and so we find
\begin{equation}
 \int_{B(\epsilon)} \left(f\imath_n A^I + \Theta^I\right)\wedge\delta\left(G_{IJ}*\upd A^J\right) = \tilde{\Phi}^{(I, \alpha)} \int_{T_\alpha(\epsilon)} \delta\left(G_{IJ}*\upd A^J\right) + \mathcal{O}(\epsilon)
\end{equation}
Here the $T_\alpha(\epsilon)$ are $(D-p-1)$-chains on $B(\epsilon)$. Finally, we use the fact that the $(D-p-1)$-chains on $B(\epsilon)$ are homologous to the $(D-p-1)$-chains on $B$ - indeed, we can find them just by translating back along the generators of the horizon. So, using the linearised field equations we have \begin{equation}
\int_{T_\alpha(\epsilon)} \delta\left(G_{IJ}*\upd A^J\right) = \int_{T_\alpha} \delta\left(G_{IJ}*\upd A^J\right) = \delta \mathcal{Q}_{(I, \alpha)}
\end{equation}
which vanishes since we are restricting to perturbations satisfying $\delta \mathcal{Q}_{(I, \alpha)} = 0$. Note that the $\tilde{\Phi}^{(I, \alpha)}$ are not related to the electrostatic potential differences of the background spacetime, as they relate to the pair $(\xi, \Theta)$ rather than the pair $(K, \Lambda)$.

Finally, we need to deal with the term
\begin{equation}
 \int_E G_{IJ}\Delta\delta A^I \wedge \imath_\xi *\upd A^J
\label{eqn3}
\end{equation}
Now, from equation (\ref{edge gauge field perturbation}) we have
\begin{equation}
 \Delta\delta A^I = (-1)^p\sum \delta \mathcal{P}^{(I, \beta)} \sigma_\beta
\end{equation}
Since the magnetic charges do not change by assumption, we have $\delta \mathcal{P}^{(I, \beta)} = 0$ and so this term also vanishes. This completes the proof of Lemma 1.

\subsection{Gauge Invariance of the Canonical Energy}
We now wish to prove that the quantity $\mathcal{E}(\delta S)$ is gauge-invariant, when $\delta S$ is axisymmetric, invariant under the action of the Killing fields on the compact manifold, and gives vanishing linearised perturbations to the ADM charges. In addition, we will require that the gauge change respects our gauge conditions at the horizon, asymptotic flatness at infinity, axisymmetry and invariance under the isometries of the compact manifold.

In particular, we will consider the transformation
\begin{equation}
 \delta S \rightarrow \delta S + \mathscr{L}_Y S + \delta_{\Theta} S
\end{equation}
where $(Y, \{\Theta^I\})$ satisfy the conditions of lemma 1, with $\Theta^I = 0$ on $B$: this is the most general gauge transformation which satisfies our gauge conditions and is also an asymptotic symmetry, i.e. it preserves the asymptotic form of the perturbation. In addition, we require $\mathscr{L}_{\psi^i}Y = \mathscr{L}_{X^j}Y = 0$. Gauge invariance under such transformations will be assured if we can show
\begin{equation}
 W_\Sigma\left(\delta S, \mathscr{L}_t(\mathscr{L}_Y S + \delta_\Theta S)\right) = 0
\label{lemma eqn1}
\end{equation}
To see that this is sufficient, let $\varphi_Y$ be the diffeomorphism generated by the Killing vector field $t$. Then, since both $S$ and $(\varphi_{Y})_* S + \delta_\Theta S$ satisfy the equations of motion and our gauge conditions, $\mathscr{L}_Y S + \delta_\Theta S$ will satisfy the linearised equations of motion and our gauge conditions. Thus the term
\begin{equation}
 W_\Sigma\left(\mathscr{L}_Y S + \delta_\Theta S, \mathscr{L}_t(\mathscr{L}_Y S + \delta_\Theta S)\right) = 0
\end{equation}
is of the same form as (\ref{lemma eqn1}), with $\delta S$ replaced by $(\mathscr{L}_Y S + \delta_\Theta S)$.

We now use
\begin{equation}
 \mathscr{L}_t\mathscr{L}_Y S + \mathscr{L}_t\delta_\Theta S = \mathscr{L}_{[t, Y]} S + \delta_{(\mathscr{L}_Y\Lambda + \mathscr{L}_t\Theta)} S
\end{equation}
From the conditions imposed above, we see that the pair $\left([t,Y], \{\mathscr{L}_Y\Lambda^I + \mathscr{L}_t\Theta^I\}\right)$ satisfies the conditions of lemma 1,and so $\mathcal{E}(\delta S)$ is gauge invariant.

\subsection{Stationary perturbations}
\label{stationary section}

At this stage there are two obstructions to the use of the canonical energy as a tool for the study of stability or instability. First, though we have shown that the canonical energy may only decrease in the future, we have not yet ruled out the possibility that it is in fact constant. Second, there are known perturbations which have $\mathcal{E}(\delta S) < 0$ and yet which do not correspond to instabilities. Indeed, \cite{Hollands:2012sf} showed that perturbations within the Schwarzschild family obey $\mathcal{E}(\delta S) < 0$. Fortunately, both problems may be overcome at once by a proper consideration of stationary perturbations.



We desire some restriction on the initial data perturbations which rules out the possibility of perturbations to other stationary spacetimes, at least when $\mathcal{E}$ takes negative values. At first sight, it may seem that this condition is already satisfied: from our definition of the canonical energy we see that $\mathcal{E} = 0$ if $\mathscr{L}_t\delta S = 0$, i.e. if $t$ is Killing in the perturbed spacetime. This interpretation would contradict our earlier statements about perturbations within the Schwarzschild family - clearly something has gone wrong.

The reason for this (apparent) contradiction is that the Killing field in the perturbed spacetime need not be the same as the Killing field in the background spacetime. In addition, we should allow for time evolution of the gauge fields, so long as they evolve by gauge transformations alone. We should therefore define a \emph{stationary perturbation} to be a perturbation such that
\begin{equation}
 \mathscr{L}_t \delta S + \mathscr{L}_{\delta t} S + \delta_\Theta S = 0
\end{equation}
We should also insist that this perturbation obeys our boundary conditions, and fall-off conditions near infinity. With these restrictions, it can be seen that pair $(-\delta t, -\Theta)$ precisely satisfy the conditions of lemma 1 above. Thus we observe that, if we restrict to stationary, axisymmetric perturbations which do not change the linearised ADM quantities, then
\begin{equation}
 \mathcal{E}(\delta S; S) = W_\Sigma(\delta S, \mathscr{L}_t \delta S;S) = W_\Sigma(\delta S, \mathscr{L}_{-\delta t}S + \delta_{-\Theta}S; S) = 0
\end{equation}
where the last equality follows from an application of lemma 1. So we see that our desired restriction is to axisymmetric perturbations which do not change the linearised ADM quantities - stationary perturbations within this class of perturbations make no contribution to the canonical energy.

This is enough to argue for instability, however, in order to make the argument for stability we really need to show that the degeneracies of $\mathcal{E}$, defined on (a dense subspace of) the space of axisymmetric perturbations which do not change the ADM quantities, are precisely the stationary perturbations. It should be possible to do this in a similar manner to \cite{Hollands:2012sf}, namely, to begin with the space of perturbations given by $(\mathscr{L}_\xi S + \delta_\Phi S)$, where $(\xi, \Theta)$ satisfy the conditions of lemma 1. We then need to show that the space of perturbations symplectically orthogonal to this\footnote{The symplectic product arises naturally when writing the perturbations in the Hamiltonian formalism.} is precisely the space of perturbations with vanishing linearised ADM quantities. It then follows immediately that, on this subspace, $\mathcal{E}$ is degenerate precisely on the stationary perturbations.

We will not go into more detail here, as the process is expected to proceed in precisely the same way as in \cite{Hollands:2012sf}. We turn instead to the final property which we require of the canonical energy: that its flux across null infinity and across the horizon is positive. Before we discuss this, we must first learn something about the rates of decay of the various fields near null infinity.

\section{Linear Stability of Asymptotic Flatness}
\label{asymptotics section}

Our aim in this section is to write down the appropriate rate of decay of the various fields and their perturbations near null infinity. Clearly, in making our definition of ``asymptotically flat'' we are free to write down any rates of decay which we desire, but the problem is to show that these decay rates are \emph{stable}, at least to linear order. By this we mean that an arbitrary (compactly supported) first order perturbation of the initial data on a spacelike hypersurface will, when allowed to propagate out to null infinity, satisfy the linearised version of our ``asymptotic flatness'' conditions. This means that linear perturbations cannot destroy our definition of asymptotic flatness, and also gives us the rate of decay of our perturbations, which we will need to use later in order to evaluate the flux of canonical energy across null infinity. See \cite{Geroch:1978ur} and \cite{Hollands:2003ie} for the vacuum case, in $D = 4$ and $D > 4$ respectively. Be aware, however, that the techniques used in those papers, as well as the one presented below, are only applicable for even $D$ - see \cite{Hollands:2004ac} for further discussion of this issue. Note that establishing the stability of an appropriate definition of asymptotic flatness is essential for our argument, since otherwise the flux of canonical energy across null infinity may not be well defined.

Note that stability in this context is not related to the actual dynamical stability of any given, asymptotically flat background solution - we are only dealing with the stability of the \emph{definition} of asymptotic flatness. In other words, on an asymptotically flat background, a compactly supported perturbation on some initial Cauchy surface will, when allowed to evolve, give rise to a new, asymptotically flat solution (at least while the linear approximation holds). This solution may differ drastically from the background solution, corresponding to a dynamical instability, but as long as the perturbations still fall off sufficiently fast near future null infinity, we say that our definition of asymptotic flatness is stable.

In this section we will assume that the scalar fields are massless, i.e. $V_{,BC} = 0$. This restriction is not essential, but the asymptotics of the scalar fields are different in the massless and massive cases. In the massless case, as we will see below, we obtain power law decay of the scalar fields, whereas in the massive case (which we will not deal with explicitly), as long as the mass is positive, we would instead obtain exponential decay. This means that the contributions from the scalar fields to the integrals at infinity which we constructed above would vanish. It is in order to see the contributions to these integrals from massless fields that we consider this case in detail below.

We will begin by fixing a gauge for our perturbations - clearly, in the presence of gauge freedom, it only makes sense to discuss rates of decay in a particular gauge. We will then make a conformal rescaling of the metric and the other fields. These conformally rescaled fields will be smooth fields on a manifold with boundary (the conformal compactification of the original manifold), and will provide us with our definition of asymptotic flatness. We will similarly conformally rescale the perturbations. We then aim to find hyperbolic equations of motion for the conformally rescaled perturbations, in which every term is smooth up to null infinity; this will prove the linear stability of asymptotic flatness.

In order to define asymptotic flatness near null infinity, we will require two additional manifolds. These are the conformally compactified, or unphysical manifold $\tilde{M}$, and the reference manifold $\bar{M}$, both of which are in fact manifolds with boundary. The metric on the unphysical manifold is related to our metric $g_{\mu\nu}$ by $\tilde{g}_{\mu\nu} = \Omega^2 g_{\mu\nu}$, where $\Omega$ is a conformal factor which vanishes at the boundary of $\tilde{M}$ and is chosen so that the unphysical metric agrees on its boundary with the metric on the boundary of a certain region of the Einstein static universe. See \cite{Hollands:2003ie} for more details on this construction. Meanwhile, the reference manifold is precisely this region of the Einstein static universe - namely, the region $\{-\pi/2 \leq t \pm \psi \leq \pi/2\}$, together with the usual metric on the Einstein static universe, $\bar{g}_{\mu\nu}$. In this section, indices on tensor fields with tildes are manipulated using the unphysical metric, while those with overbars are manipulated with the reference metric, etc.

Our first criterion for asymptotic flatness is the presence of diffeomorphisms between the ``asymptotic regions'' of these three manifolds. The asymptotic regions are the complements in each manifold of some compact region. These allow us to identify tensor fields (in the asymptotic region) on $M$ with tensor fields on $\tilde{M}$ and $\bar{M}$, and in the following statements this identification is implicit. Given a conformal factor $\Omega$, a tensor field $T_{\mu\nu\ldots}$ is said to be $\mathcal{O}(\Omega^s)$ if
$\Omega^{-s}T_{\mu\nu\ldots}$ is smooth at the boundary of $\bar{M}$. We will also make use of the covector $n_\mu = \partial_\mu \Omega$ in this section, which should not be confused with the null generator of the horizon, which will not arise in this section.

When making comparisons between the various metrics, the gauge freedom in our metric translates into the freedom in choosing the diffeomorphisms between the manifolds. There is additional gauge freedom in the gauge
fields, however, we are able to make our definition of asymptotic flatness relative to the field strengths $\upd A^I$ rather than the gauge fields themselves. We are now ready to make our definition of asymptotic flatness near null infinity:\\[10pt] {\bf Definition 1:} A solution to the equations of motion arising from the Lagrangian (\ref{lagrangian}) is said to be \emph{weakly asymptotically simple} at null infinity if there exists a choice of conformal factor $\Omega$, and a choice of diffeomorphisms between the appropriate regions of the manifolds $M$, $\tilde{M}$ and $\bar{M}$ defined above, such that
\begin{equation}
 \begin{split}
  \bar{g}_{\mu\nu} - \tilde{g}_{\mu\nu} &= \mathcal{O}\left(\Omega^{\frac{D-2}{2}}\right)\\
  \bar{\epsilon}_{\mu_1 \ldots \mu_D} - \tilde{\epsilon}_{\mu_1 \ldots \mu_D} &= \mathcal{O}\left(\Omega^\frac{D}{2}\right) \\
  \left(\bar{g}^{\mu\nu} - \tilde{g}^{\mu\nu}\right)n_\mu &= \mathcal{O}\left(\Omega^{\frac{D}{2}}\right)\\
  \left(\bar{g}^{\mu\nu} - \tilde{g}^{\mu\nu}\right)n_\mu n_\nu &= \mathcal{O}\left(\Omega^\frac{D+2}{2}\right)\\
  \phi^B &= \mathcal{O}\left(\Omega^{\frac{D-2}{2}}\right)\\
  \upd A^I &= \mathcal{O}\left(\Omega^{\frac{D-4 - 2p}{2}}\right)\\
  \imath_n  \upd A^I &= \mathcal{O}\left(\Omega^{\frac{D-6 - 2p}{2}}\right)
 \end{split}
\label{asympflat}
\end{equation}
\\[10pt] We now need to show that this definition is linearly stable to perturbations, in the sense described above. We will first impose some gauge conditions on our perturbations, since otherwise we would have no hope of proving decay results for the perturbations. In the vacuum case\footnote{in $D > 4$ - a different gauge can be used for the $D=4$ case \cite{Geroch:1978ur}} the transverse, traceless gauge can be used, but it is not possible to impose the traceless condition in our case. Instead, we impose a modified transverse gauge, together with the Lorentz gauge choice for the gauge fields:
\begin{equation}
 \begin{split}
  \nabla^\nu \delta g_{\mu\nu} - \frac{1}{2}\nabla_\mu \delta g - \Omega^{-1}n_\mu \delta g &= 0 \\
  \nabla^\nu \delta A_{\nu\mu_1 \ldots \mu_{p-1}} &= 0\\
 \end{split}
\label{asymp gauge}
\end{equation}

There are two points to note about equation (\ref{asymp gauge}). First, the presence of the term $\Omega^{-1}n_\mu \delta g$ may give cause for concern, since it appears to diverge in the asymptotic region. However, it is shown in appendix B that this gauge may be imposed upon asymptotically flat initial data for the linearised equations.

Secondly, it should be noted that we have not completely fixed the gauge, either for the metric perturbation or for the gauge field perturbations. We can contrast our metric gauge condition with the usual transverse, traceless condition used in \cite{Hollands:2003ie} - clearly, a scalar gauge degree of freedom remains unfixed in our gauge, roughly corresponding to the trace of the metric perturbation. However, this degree of freedom propagates according to a scalar wave equation - this is precisely the reason why it can be fixed to zero for all time in vacuum. As such, its asymptotics are dictated by this wave equation - and so, for our purposes, we do not \emph{have} to fix it, as it will already decay appropriately. A similar situation arises for the gauge fields, where we would like to be able to fix the longitudinal gauge, $\imath_n \delta A^I = 0$. This can be done in Minkowski space, but not in the presence of curvature, so again we are left with unfixed gauge degrees of freedom, but these gauge degrees of freedom admit a well-posed Cauchy problem, and obey an equation of motion which leads to appropriate asymptotics. For further details, see appendix B.

Next, we need to make a choice of variables, which will be related to appropriately conformally rescaled combinations of our perturbations. We choose the following set of variables:
\begin{equation}
\begin{split}
 \tau_{\mu\nu} &= \Omega^{-\frac{D-6}{2}}\delta g_{\mu\nu} \\
 \hat{\tau}_\mu &= \Omega^{-\frac{D-4}{2}}\tilde{n}^\nu\delta g_{\mu\nu} \\
 \tilde{\tau} &= \Omega^{-\frac{D-4}{2}}\tilde{g}^{\mu\nu}\delta g_{\mu\nu} \\
 \sigma &= \tilde{\nabla}^\mu\tau_\mu \\
 \tilde{\phi}^B &= \Omega^{-\frac{D-2}{2}}\delta\phi^B \\
 \delta \tilde{A}^I &= \Omega^{-\frac{D-2}{2}+p}\delta A^I \\
 \delta \hat{A}^I &= \Omega^{-\frac{D-2}{2}+p}\imath_{\tilde{n}}\delta A^I \\
\end{split}
\label{null variables}
\end{equation}
Note that the conditions that these variables are smooth up to the boundary of $\tilde{M}$ are precisely the linearisations of our asymptotic flatness conditions (\ref{asympflat}).

If we denote the set of variables above by $\{q^N\}$, then we are faced with the task of finding equations of the form
\begin{equation}
 P^{\mu\nu}_{M,N}\tilde{\nabla}_\mu \tilde{\nabla}_\nu q^N = f^N(q^M, \tilde{\nabla} q^M)
\end{equation}
where the functions $f^N$ are explicitly smooth up to the boundary, i.e. they cannot contain any negative powers of $\Omega$. Moreover, we require this system to be (strictly) hyperbolic, which is the condition that there exists some covector $\tau$ such that the polynomial
\begin{equation}
P(s;\xi) = \det{(P^{\mu\nu}_{M,N}(\xi_\mu + s\tau_\mu)(\xi_\nu + s\tau_\nu))} \\
\end{equation}
has $D$ distinct roots for all $\xi \neq 0$.

This is done in detail in appendix A. Despite the complexity of these equations, the result is the expected one: the conditions on the metric for asymptotic flatness are the same as in the vacuum case, while the conditions on the fields are the same as the conditions usually imposed in Minkowski space. This is not surprising, as terms arising from couplings between the various fields tend to fall off much faster than the fields themselves, since they include (by definition) multiple fields, all of which are decaying near null infinity. Only the terms in the equations of motion which are linear in all fields are important, and these do not include the coupling terms.

\subsection{Discussion}
\label{asymptotics subsection}
There are some issues with this approach. The machinery of null infinity described above exists only in even spacetime dimensions (see \cite{Hollands:2004ac}), although with appropriate definitions in place this may possibly be overcome. The reason for this difficulty is that half-integer powers of $\Omega$ appear in odd dimensions, and these spoil various smoothness assumptions. In addition, we have had to restrict to perturbations which are compactly supported on the initial, spacelike slice, whereas we are actually interested in asymptotically flat perturbations. These are defined as perturbations on an initial data slice through spacelike infinity such that the variables defined in (\ref{null variables}) are smooth at spacelike infinity, and where, \emph{in addition} the derivatives of these variables in the $\tilde{n}$ direction fall off one power of $\Omega$ faster. This additional restriction on the initial data is the reason why the additional ``mass terms'' in section \ref{first law subsection}, equations (\ref{scalarcontribution}) and (\ref{gauge field contribution}), vanish. However, our energy functional is continuous (in the sense of $L^2$) and the compactly supported perturbations are expected to be dense in the set of asymptotically flat perturbations with vanishing linearised ADM quantities (this was proved for gravitational perturbations in \cite{Hollands:2012sf}).

Finally, the gauge choice and choice of variables outlined above fail to provide a set of hyperbolic equations of motion in four dimensions, the reason being that a factor of $(D-4)$ appears in front of the Laplace-Beltrami operator, spoiling hyperbolicity. Note that this also occurs when using the transverse, traceless gauge in the vacuum case; an alternative choice of gauge and of variables was found in \cite{Geroch:1978ur} which works in four dimensions, and it is likely that something similar can be achieved in our case.

Finally, we mention that the definitions and proofs of linear stability carry over straightforwardly to the asymptotically Kaluza Klein case, where we say that a spacetime is asymptotically Kaluza Klein if, outside of some compact region, it is diffeomorphic to the product of an asymptotically flat spacetime and some (fixed) compact manifold, and moreover, where the metric on the compact manifold approaches some fixed (i.e. independent of the position on the asymptotically flat manifold) metric at an appropriate rate near infinity. Then the ``$D$'' appearing in the formulae above must be replaced by the dimension of the asymptotically flat space, rather than the full dimension of the space.

\section{Flux of Canonical Energy Across the Horizon and Null Infinity}

\subsection{Flux Across Null Infinity}
Our asymptotic flatness conditions mean that, close to null infinity, we can define a function $u$ and work in a gauge such that the unphysical line element, associated with $\tilde{g}_{\mu\nu}$ becomes \begin{equation}
 \upd s^2 = 2\upd u\upd \Omega + \tilde{\mu}_{ij} \upd x^i \upd x^j + \mathcal{O}(\Omega)
\label{asymp metric}
\end{equation}
where $\tilde{\mu}_{ij}$ is a Riemannian metric on some compact $(d-2)$ dimensional manifold, with associated volume form $\tilde{\epsilon}_{\tilde{\mu}}$. Then $\tilde{n}^\mu = (\partial/\partial u)^\mu$ is proportional, at null infinity, to the asymptotically timelike Killing vector fields via the formula \begin{equation}
 \tilde{t}^\mu = (\tilde{t}^\nu \partial_\nu u)\tilde{n}^\mu
\end{equation}
where $\tilde{t}^\nu \partial_\nu u > 0$ is constant on null infinity.

Using this, together with our decay properties and gauge conditions, we can calculate the pull-back of the symplectic current to null infinity. The gravitational contributions are found in \cite{Hollands:2012sf} to be
\begin{equation}
 \omega_{\text{GR}}(\delta g, \mathscr{L}_t \delta g) = 2(\tilde{t}^\mu\partial_\mu u)\delta \tilde{N}_{ij}\delta \tilde{N}^{ij}\upd u\wedge\tilde{\epsilon}_{\mu} + (\tilde{t}^\mu\partial_\mu u)\upd\left(\tilde{\tau}^{\mu\nu}(\mathscr{L}_{\tilde{n}}\tilde{\tau}_{\mu\nu})\tilde{\epsilon}_\mu\right)
\end{equation}
where $N_{ij}$ is the Bondi news tensor. Note that the second term gives a boundary term, while the first is manifestly positive, as the contraction is taken with respect to $\tilde{\mu}_{ij}$.

We now examine the additional contributions arising from the scalar fields and gauge fields. A short calculation shows that these are given respectively by
\begin{equation}
 \begin{split}
  \omega_{\text{scalar}}(\delta S, \mathscr{L}_t \delta S) &= 2(\tilde{t}^\mu\partial_\mu u)F_{BC}(\mathscr{L}_{\tilde{n}}\tilde{\phi}^B)(\mathscr{L}_{\tilde{n}}\tilde{\phi}^C)\upd u\wedge\tilde{\epsilon}_\mu + \upd\left(F_{BC}\tilde{\phi}^B\mathscr{L}_{\tilde{n}}\tilde{\phi}^C \tilde{\epsilon}_\mu\right) \\
  \omega_{\text{gauge}}(\delta S, \mathscr{L}_t \delta S) &= 2(\tilde{t}^\mu\partial_\mu u)G_{IJ}(\mathscr{L}_{\tilde{n}}\tilde{A}^I)^{i_1 \ldots i_p}(\mathscr{L}_{\tilde{n}}\tilde{A}^J)_{i_1 \ldots i_p}\upd u\wedge\tilde{\epsilon}_\mu + \upd\left(G_{IJ}(\tilde{A}^I)^{i_1 \ldots i_p}\mathscr{L}_{\tilde{n}}(\tilde{A}^J)_{i_1 \ldots i_p} \tilde{\epsilon}_\mu\right)
 \end{split}
\end{equation}
where the indices $\{i_1 , \ldots i_p\}$ represent the components in the $x^{i_1}, \ldots x^{i_p}$ directions, which are contracted using the metric $\tilde{\mu}_{ij}$. Note that, because all the background fields fall off near infinity (except $\tilde{g}_{\mu\nu}$, which is $\mathcal{O}(1)$), the only terms which survive near infinity are the ones mentioned above. In particular, $\omega_{\text{scalar}}$ is independent of $\delta g_{\mu\nu}$ and $\delta A^I$, etc.

Finally, we remark that all of the above expressions are sums of positive definite quantities and boundary terms. Thus , if $\mathscr{I}_{12}$ is a section of null infinity, with $\partial\mathscr{I}_{12} = \mathscr{C}(u_2) - \mathscr{C}(u_1)$ , then we have
\begin{equation}
 \begin{split}
 W_{\mathscr{I}_{12}}(\delta S, \mathscr{L}_t \delta S) &= 2\int_{\mathscr{I}_{12}}(\tilde{t}^\mu \partial_\mu u)\left(\delta\tilde{N}^{\nu\rho}\delta\tilde{N}_{\nu\rho} + F_{BC}\mathscr{L}_{\tilde{n}}\varphi^B\mathscr{L}_{\tilde{n}}\varphi^C + \right. \\
 &\phantom{AAA} \left. G_{IJ}(\mathscr{L}_{\tilde{n}}\tilde{\alpha}^I)^{i_1 \ldots i_p}(\mathscr{L}_{\tilde{n}}\tilde{\alpha}^J)_{i_1 \ldots i_p}\vphantom{\tilde{N}^{\nu\rho}}\right)\upd u\wedge\tilde{\epsilon}_\mu + C(u_1) - C(u_1)
 \end{split}
\end{equation}
where the boundary terms $C_i$ are given by
\begin{equation}
 C(u_i) = \int_{\mathscr{C}(u_i)} (\tilde{t}^\mu \partial_\mu u)\left(\tilde{\tau}^{\mu\nu}\mathscr{L}_{\tilde{n}}\tau_{\mu\nu} + F_{BC}\varphi^B\mathscr{L}_{\tilde{n}}\varphi^C + G_{IJ}(\tilde{\alpha}^I)^{i_1 \ldots i_p}\mathscr{L}_{\tilde{n}}(\tilde{\alpha}^J)_{i_1 \ldots i_p}\right)\tilde{\epsilon}_\mu
\end{equation}

\subsection{Flux Across the Horizon}

We now aim to find a similar formula for the flux of canonical energy across the horizon. The process is very similar to that described above - note the similarity between (\ref{gaussian}) and (\ref{asymp metric}). In this case, however, the computation is a little more involved, since we cannot appeal to fall off rates in order to ``decouple'' the various parts of the symplectic current, as was done above. Instead, we must appeal directly to our gauge and boundary conditions. Nevertheless, if $\mathscr{H}_{12}$ is a section of the future horizon, the boundary conditions together with axisymmetry and the stationarity of the background, and our gauge condition, $\imath_n\delta A^I$ on $\mathscr{H}$, are sufficient to ensure that
\begin{equation}
\begin{split}
 W_{\mathscr{H}_{12}}(\delta S, \mathscr{L}_t \delta S) &= \int_{\mathscr{H}_{12}}2(\kappa u)\left(4\delta \sigma_{\alpha \beta}\delta\sigma^{\alpha\beta} + F_{BC}\mathscr{L}_n \delta\phi^B \mathscr{L}_n \delta\phi^C \right. \\
 &\phantom{AAA} \left.+ G_{IJ}(\mathscr{L}_n \delta A^I)^{\alpha_1 \ldots \alpha_p}(\mathscr{L}_n \delta A^J)_{\alpha_1 \ldots \alpha_p} \right)\upd u\wedge\epsilon_\mu + B(u_2) - B(u_1)
\end{split}
\end{equation}
where $\kappa > 0$ is the surface gravity, $\delta \sigma_{\alpha\beta}$ is the perturbed shear of the horizon, the indices $\alpha, \beta$ refer to the components with respect to the basis of one forms $\{\gamma^\alpha\}$ which are raised using the (Riemannian) metric $\mu^{\alpha\beta}$ (see (\ref{gaussian})), and $\epsilon_{\mu}$ is the induced metric on the intersection of the future horizon with surfaces of constant $u$. Note also that $u$ and $n$ above are associated with the horizon, as in (\ref{gaussian}) and the preceding discussion, rather than with null infinity as in (\ref{asymp metric}). Finally, if $\partial\mathscr{H}_{12} = \mathscr{B}(u_2) - \mathscr{B}(u_1)$ then the boundary terms $B(u_1)$ and $B(u_2)$ are given by
\begin{equation}
 B(u_i) = \int_{\mathscr{B}(u_i)}(\kappa u_i)\left(\delta g^{\mu\nu}(\mathscr{L}_n \delta g)_{\mu\nu} + F_{BC}\delta\phi^B \mathscr{L}_n \delta\phi^C + G_{IJ}(\delta A^I)^{\alpha_1 \ldots \alpha_p}(\mathscr{L}_n \delta A^J)_{\alpha_1 \ldots \alpha_p} \right)\epsilon_\mu
\label{horizon boundary term}
\end{equation}

To find the gravitational terms in the above formulae it is essential first to show that the perturbed expansion of the horizon, $\delta\vartheta$ vanishes everywhere along the future horizon, rather than just at the bifurcation surface (see \cite{Hollands:2012sf} for details). The Raychaudhuri equation on $\mathscr{H}_+$ is
\begin{equation}
 \frac{\upd}{\upd u}\vartheta(\lambda) = -\frac{1}{D-2}\vartheta(\lambda)^2 - \sigma_{\alpha\beta}\sigma^{\alpha\beta} - T_{\mu\nu}(\lambda)n^\mu n^\nu
\end{equation}
We know that in the background spacetime, $\vartheta(0) = \sigma_{\alpha\beta}(0) = 0$, and that $n$ is null in both the background and perturbed spacetimes. 
In addition, the background Raychaudhuri equation implies that $\imath_n \upd A^I = \upd r \wedge A^I_{(r)}$ on the horizon, for some $(p-1)$-form $A^I_{(r)}$ - see (\ref{horizon energy flux}). Using these facts we see that $\upd \delta\vartheta/\upd u = 0$. Since we choose $\delta\vartheta = 0$ at $u = 0$, this implies that $\delta\vartheta = 0$ along the whole future horizon, as required.

\subsection{Modified Canonical Energy}
We now define the \emph{modified canonical energy} $\bar{\mathcal{E}}$, evaluated on a surface $\Sigma$ with one boundary, $\mathscr{C}(u)$, at null infinity and an interior boundary, $\mathscr{B}(u)$, on the future horizon, by
\begin{equation}
 \bar{\mathcal{E}}(\delta S, u) = \mathcal{E}(\delta S) - C(u) - B(u)
\end{equation}
This modified canonical energy has the desired property that it is non-increasing, in the sense that, if one such surface $\Sigma_1$ lies entirely within the causal past of another such surface $\Sigma_2$, then
\begin{equation}
 \bar{\mathcal{E}}_{\Sigma_2} \leq \bar{\mathcal{E}}_{\Sigma_1}
\end{equation}
This follows straightforwardly from the fact that the symplectic form is closed, which means that
\begin{equation}
 \mathcal{E}_{\Sigma_2} = \mathcal{E}_{\Sigma_1} + W_{\mathscr{I}_{12}}(\delta S, \mathscr{L}_t \delta S) + W_{\mathscr{H}_{12}}(\delta S, \mathscr{L}_t \delta S)
\end{equation}
then the flux terms $ W_{\mathscr{I}_{12}}(\delta S, \mathscr{L}_t \delta S)$ and $W_{\mathscr{H}_{12}}(\delta S, \mathscr{L}_t \delta S)$ are the sum of positive definite terms and boundary terms, which are absorbed into the definition of the modified canonical energy.

Note that the modified canonical energy agrees with the canonical energy as the boundaries approach spatial infinity and the bifurcation surface respectively, since the Lie derivatives of our perturbations in the $\tilde{n}$ direction tend to zero faster on this slice\footnote{This is also manifestly true for compactly supported perturbations} (see section \ref{asymptotics subsection}), and the $u_i$ appearing in (\ref{horizon boundary term}) vanishes, by definition, at the bifurcation surface. In addition, if the perturbation approaches some stationary configuration in the far future, the boundary terms should also vanish and so the modified canonical energy should approach the canonical energy.

Note that we have finally proved that the (modified) canonical energy possesses all the properties enumerated at the start of subsection $6$.

\section{Stability and Instability}

We can now give arguments relating the positivity properties of $\mathcal{E}$ to the stability properties of the background solution. We can provide fairly strong arguments for instability in the case where a perturbation can be found for which $\mathcal{E}(\delta S)$ is negative. On the other hand, there are a number of limitations on the statements which can be made about stability in the case when $\mathcal{E} > 0$ can be established for all permissible perturbations.

\subsection{Instability}
Suppose that an asymptotically flat, axisymmetric perturbation can be found with vanishing linearised ADM quantities and which satisfies $\mathcal{E}(\delta S) < 0$. Suppose in addition that this perturbation approaches a stationary solution in the far future. Then we must have $\bar{\mathcal{E}}(\delta S) \rightarrow \mathcal{E}(\delta S_0)$ for some stationary perturbation of the original solution, $\delta S_0$, which must also have vanishing linearised ADM quantities\footnote{Recall that, at linear order, neither mass nor angular momenta can be radiated. In the theory under consideration, neither can charge, as there are no charged fields.}. However, as shown in section \ref{stationary section}, such a perturbation has vanishing canonical energy, so $\bar{\mathcal{E}}(\delta S) \rightarrow 0$. On the other hand, the modified canonical energy cannot increase in the future, and is negative initially, giving a contradiction.

We interpret this as evidence for instability. Although this does not prove that there is some physical quantity which grows with time (let alone exponentially with time), as long as the linear approximation is valid, we have shown that the perturbation cannot decay, nor can it approach a stationary configuration. There are only two remaining options: either the perturbation grows large enough for the linear approximation to become invalid (signalling an instability), or the perturbation remains small without ever becoming stationary, nor radiating to null infinity (since radiation decreases the modified canonical energy), nor falling through the black hole horizon. This latter case appears unlikely, even in a theory as general as the one we have considered, as non-stationary configurations are always expected to radiate in an asymptotically flat spacetime.

Note that, even if this case can be ruled out, we still cannot say anything about the rate of growth of instabilities found using this method. This may be seen as a disadvantage, although it also indicates that this approach is sensitive to instabilities which grow, say, polynomially\footnote{We would expect linear growth to correspond to perturbations with vanishing canonical energy} with time, rather than exponentially, as is usually the case.

\subsection{Stability}
The arguments for stability are a lot weaker in nature than those for instability. Suppose that, for a given background solution, we can show that $\mathcal{E}$ is positive semi-definite when restricted to axisymmetric perturbations which do not change the linearised ADM charges. As we expect the degeneracies of $\mathcal{E}$ to be given precisely by the stationary perturbations, we can form a new space of perturbations by ``quotienting out'' by stationary perturbations. Then $\mathcal{E}$ provides a non-increasing norm on this space, meaning that the canonical energy can be used to bound the ``size'' of the equivalence class of a perturbation as it evolves in time. Since the perturbation can be expressed as the sum of a stationary perturbation and a representative element of the equivalence class, and the former does not evolve while the latter is bounded, this establishes boundedness of the perturbation. We should note that this approach can only establish boundedness and not decay of the perturbation. Also, we note that in simple cases (such as four dimensional Einstein-Maxwell theory) there are no stationary perturbations about a stationary background which leave the linearised ADM charges unchanged, as the uniqueness theorems allow us to label all stationary spacetimes by their ADM charges. In such cases we do not have to take the quotient described above.

There are some issues with the argument above. First, we have had to restrict to perturbations which do not change the linearised ADM charges. This is not expected to cause many problems as we have restricted consideration to the non-extremal case, where we expect the presence of families of stationary solutions labelled by their ADM charges. Then an arbitrary perturbation could be expressed as the sum of a perturbation within this family (which is evidently stable), and a perturbation satisfying our previous conditions. Note that different arguments have recently been given, which suggest that extremal black holes are generically unstable \cite{Aretakis:2012ei} \cite{Lucietti:2012sf}.

Finally, note that we can only conclude boundedness for axisymmetric perturbations, for reasons detailed throughout this paper. Essentially, the problem is with the possible existence of an ergosphere, which would spoil positivity of energy flux through the horizon - this is connected with the phenomena of superradiance. As is well known, axisymmetric fields do not exhibit superradiance, so we have avoided dealing with this difficulty in the approach detailed above. Another possible approach is to use a vector field other than the asymptotically timelike Killing vector field $t$ in the definition of the canonical energy (\ref{canonicaldefinition2}). As was noted in \cite{Hollands:2012sf}, using the horizon Killing field $K$ results in positive flux over the horizon for arbitrary perturbations, but will not lead to positive energy flux across null infinity. We could look for a vector field which interpolates between $t$ near infinity and $K$ near the horizon, which would lead to positive energy flux across both boundaries. However, such a vector field would not be Killing, leading to a nonzero ``bulk'' contribution to the energy. In order to establish stability for arbitrary perturbations, we would need also to establish the positivity of this bulk contribution for some choice of vector field, or at least give a sufficiently stringent lower bound on its size. This is very similar to the \emph{redshift vector field} approach used successfully in studies of linear waves on black hole backgrounds \cite{Dafermos:2005eh}.

\section{An Application: The Gubser-Mitra Conjecture}

As was previously done in the vacuum case \cite{Hollands:2012sf}, we now seek to apply these methods to the \emph{Gubser-Mitra conjecture} \cite{Gubser:2000ec}. Put simply, this states that a black string (or more generally a black membrane) which is \emph{thermodynamically unstable} is classically unstable, if the circle (or torus) around which the black string is wrapped is large enough. We will consider uniform black membranes - that is, solutions to our Lagrangian (\ref{lagrangian}) which are the warped product of an asymptotically flat manifold and a $d$-dimensional torus. We will label co-ordinates for the torus as $\{z^m\}$, and insist that each of the $\partial/\partial z^m$ is a Killing vector field\footnote{Note that we do not specify that the line element on the torus is simply $\upd z^m \upd z^n \delta_{mn}$ - indeed, for charged black strings this is not the case.}. We also assume that the solutions have vanishing linear momentum, both in the asymptotically flat directions and in the compact $z^m$ directions, and that all tensors are ``block diagonal'', in the sense that their components in the mixed toroidal/asymptotically flat directions vanish. To define thermodynamic stability, suppose that there is a family of such solutions, labelled by their ADM charges. The horizon area of members of this family can be thought of as a function of the ADM charges. The Hessian of the horizon area is then \begin{equation}
 \bm{H} = \left(\begin{array}{cccc} \frac{\partial^2 \mathcal{A}}{\partial M^2} & \frac{\partial^2 \mathcal{A}}{\partial \mathcal{Q}_{(I, \alpha)} \partial M} & \frac{\partial^2 \mathcal{A}}{\partial \mathcal{P}^{(J, \beta)} \partial M} & \frac{\partial^2 \mathcal{A}}{\partial J_i \partial M} \\
 \frac{\partial^2 \mathcal{A}}{\partial M \partial \mathcal{Q}_{(K, \gamma)}} & \frac{\partial^2 \mathcal{A}}{\partial \mathcal{Q}_{(I, \alpha)}\mathcal{Q}_{(K, \gamma)}} & \frac{\partial^2 \mathcal{A}}{\partial \mathcal{P}^{(J, \beta)} \partial \mathcal{Q}_{(K, \gamma)}} & \frac{\partial^2 \mathcal{A}}{\partial J_i \partial \mathcal{Q}_{(K, \gamma)}} \\
 \frac{\partial^2 \mathcal{A}}{\partial M \partial \mathcal{P}^{(L, \delta)}} & \frac{\partial^2 \mathcal{A}}{\partial \mathcal{Q}_{(I, \alpha)} \mathcal{P}^{(L, \delta)}} & \frac{\partial^2 \mathcal{A}}{\partial\mathcal{P}^{(J, \beta)}\partial\mathcal{P}^{(L, \delta)}} & \frac{\partial^2 \mathcal{A}}{\partial\mathcal{P}^{(L, \delta)}\partial J_i}\\ \frac{\partial^2 \mathcal{A}}{\partial M \partial J_j} & \frac{\partial^2 S}{\partial \mathcal{Q}_{(I, \alpha)} \partial J_j} & \frac{\partial^2 \mathcal{A}}{\partial \mathcal{P}^{(J, \beta)}\partial J_j} & \frac{\partial^2 \mathcal{A}}{\partial J_i \partial J_j}
\end{array}\right)
\label{hessian}
\end{equation}
We say that a solution is thermodynamically unstable if $\bm{H}$ has a positive eigenvalue. Note that, with this definition, we are not considering solutions with nonzero Kaluza Klein momenta within our ensemble.

Let $\bm{\xi}$ and $\bm{v}$ be arbitrary vectors in the parameter space for this family (i.e. the space of ADM charges, excluding the linear momenta), and consider the one-parameter family of solutions
\begin{equation}
S(\lambda) = S(\bm{\xi} + \lambda\bm{v})
\end{equation}
From (\ref{secondvariation}) we see that, for this perturbation
\begin{equation}
 \mathcal{E} = -\kappa\frac{\upd^2 \mathcal{A}}{\upd \lambda^2}(0)
\end{equation}
but also
\begin{equation}
 \frac{\upd^2 \mathcal{A}}{\upd \lambda^2}(0) = \bm{v}^T\bm{H}(\bm{\xi})\bm{v}
\end{equation}
Thus, if $S(\bm{\xi})$ is thermodynamically unstable, then we are able to find a perturbation making $\mathcal{E}$ negative.

However, we cannot yet deploy the machinery developed earlier, as this perturbation clearly does not have vanishing linearised ADM charges. Our approach will be to multiply the initial data for our perturbation by a factor which is periodic on the torus, in order to achieve vanishing linearised ADM charges. However, this initial data will no longer satisfy the linearised constraints, and so additional terms are required to correct this. Finally, we will need to show that the contributions of all these additional terms to the canonical energy vanish in the limit where the size of the torus tends to infinity.

In this section we will use the notation natural to the Hamiltonian formulation, the details of which are given in appendix B. The initial data for $S(\bm{\xi})$ is given by $\left(h, \pi, \phi^B, p_B, A^I, \Pi_I\right)$, where $h_{ab}$ is the induced metric on a spacelike hypersurface, the $\phi^B$ are the restrictions of the scalar fields to this surface, and the $A^I$ are the pull backs to this surface of the gauge fields. The momenta canonically conjugate to these variables are $\pi^{ab}$, $p_B$ and $\Pi_I$ respectively. Our original perturbation, $\partial S(\lambda)/\partial \lambda$, induces initial data for the linearised equations $\left(\delta\hat{h}, \delta\hat{\pi}, \delta\hat{\phi}^B, \delta\hat{p}_B, \delta\hat{A}^I, \delta\hat{\Pi}_I\right)$, which satisfies the constraint equations. We assume that our Cauchy surface is chosen to be maximal in both the background and the perturbed spacetimes (see \cite{Chrusciel:1993cv}), so that
\begin{equation}
 \pi^{ab}h_{ab} = \delta\hat{\pi}^{ab}h_{ab} + \pi^{ab}\delta\hat{h}_{ab} = 0
\end{equation}
We also assume, for notational simplicity, that all the circles with tangents $\partial/\partial z^i$ have the same ``length'', which we denote by $\ell$, so that $z^i \sim z^i + \ell$.

The perturbations we will use are:
\begin{equation}
 \begin{split}
  \delta h_{ab} &= \left(\delta\hat{h}_{ab} + \psi h_{ab}\right)e^{ik\cdot x/\ell} \\
  \delta \pi^{ab} &= \left(\delta\hat{\pi}^{ab} - \psi\pi^{ab}\right)e^{ik\cdot x/\ell} + \sqrt{h}\mathcal{D}^{(a}\hat{X}^{b)} - \sqrt{h}\frac{1}{D}\mathcal{D}^c\hat{X}_c h^{ab}\\
  \delta A^I &= \delta\hat{A}^Ie^{ik\cdot x/\ell}\\
  \delta\Pi_I &= \delta\hat{\Pi}_Ie^{ik\cdot x/\ell}\\
  \delta\phi^B &= \delta\hat{\phi}^Be^{ik\cdot x/\ell}\\
  \delta p_B &= \delta\hat{p}_Be^{ik\cdot x/\ell}
 \end{split}
\label{GM perts}
\end{equation}
In the above, $k\cdot x = k_iz^i$ for integer constants $k_i$, and the scalar field $\psi$ and vector field $\hat{X}^a$ are to be determined. We write $k^a$ for the vector field $k_i h^{ij} (\partial/\partial z^j)^a$, and $k^i = h^{ij}k_j$ (which are not necessarily constants, unlike the $k_i$, although they are still independent of the $z^i$).

We decompose our vector field $\hat{X}^a$ as
\begin{equation}
 \hat{X}^a = \left(X^a + \zeta k^a\right)e^{ik\cdot x/\ell}
\end{equation}
where $X^a$ is orthogonal to the surfaces of constant $z^i$. In order to ensure that our perturbations leave the linearised ADM charges invariant, we will also need to demand that the Lie derivatives in the $z^i$ directions of the scalar fields $\psi$ and $\zeta$ and of the vector field $X^a$ vanish.

The linearised Gauss Law constraint is automatically satisfied, since the extra contribution from the periodic factor we have inserted is
\begin{equation}
 k^a (\Pi^I)_{a b_1 \ldots b_{p-1}}
\end{equation}
but the momenta in the internal ($z^i$) directions vanish by assumption. The linearised momentum constraint implies
\begin{equation}
 -\mathcal{D}^b \mathcal{D}_b \hat{X}_a - \mathcal{D}_b \mathcal{D}_a \hat{X}^b + \frac{2}{D}\mathcal{D}_a \mathcal{D}_b \hat{X}^b = \frac{i}{\ell\sqrt{h}}\left(\pi^{bc}\delta \hat{h}_{bc} + p_A \delta\hat{\phi}^A + (\Pi^I)^{a_1 \ldots a_p}(\delta\hat{A}_I)_{a_1 \ldots a_p}\right)e^{ik\cdot x/\ell} k_a
\end{equation}
Note that we should be careful to distinguish between the components of this vector field in the $z^i$ direction and the functions $k^i$ defined above. 

By decomposing $\hat{X}$ in the way we have indicated above, the momentum constraint may be separated into two - one which points in the $z^i$ directions, and one which points entirely in the ``external'' directions, i.e the directions in the surfaces of constant $z^i$. Since nothing depends on the $z^i$, each of these may then be written entirely as equations on some surface of constant $z^i$. We use indices from the start of the Greek alphabet to denote quantities and derivatives induced in the``external'' directions (i.e. on the constant $z^i$ planes), while indices from the middle of the Latin alphabet label components in the ``internal'' directions. We emphasise that, in the equations below, quantities labelled by internal indices are to be viewed as scalar functions on our surface of constant $z^i$, with derivatives being taken accordingly. In addition, all derivatives are to be taken with respect to the induced metric on the surface of constant $z^i$, rather than the full metric. Finally, we define $\mathsf{h} = \det(h_{ij})$. Then the two equations are
\begin{equation}
 \begin{split}
  0 &= \mathcal{D}_\beta\left(\sqrt{\mathsf{h}}\mathcal{D}^\beta X^\alpha\right) - \frac{(D-2)}{D}\mathcal{D}_\beta\left(\sqrt{\mathsf{h}}\mathcal{D}^\alpha X^\beta\right) - \frac{(D-4)}{4D}\sqrt{\mathsf{h}}h^{ij}h^{kl}(\partial^\alpha h_{ik})(\partial_\beta h_{jl})X^\beta \\
  & \phantom{A} - \frac{(D-2)}{2D}\sqrt{\mathsf{h}}h^{ij}\left(\mathcal{D}^\alpha \partial_\beta h_{ij}\right)X^\beta - \sqrt{\mathsf{h}}\frac{k^2}{\ell^2}X^\alpha - \frac{2(D-1)}{D}\sqrt{\mathsf{h}}R^{\alpha\beta}X_\beta - \frac{i}{\ell}\frac{(D-2)}{D}\sqrt{\mathsf{h}}\partial^\alpha(k^2 \zeta) \\
  & \phantom{A} + \frac{i}{\ell}\sqrt{\mathsf{h}}k^ik^j(\partial^\alpha h_{ij})\zeta
 \end{split}
\label{X equation}
\end{equation}
and
\begin{equation}
 \begin{split}
  & \mathcal{D}_\beta\left(k^2 \sqrt{\mathsf{h}}\partial^\beta \zeta\right) - \frac{1}{2}\sqrt{\mathsf{h}}k^ik^j\left(\mathcal{D}^\beta\partial_\beta h_{ij}\right)\zeta + \sqrt{\mathsf{h}}h^{ij}k^kk^l(\partial^\beta h_{ij})(\partial_\beta h_{jl})\zeta \\
  & \phantom{A} - \frac{1}{4}\sqrt{\mathsf{h}}h^{ij}k^kk^l(\partial^\beta h_{ij})(\partial_\beta h_{kl})\zeta - \frac{2}{D}\sqrt{\mathsf{h}}\frac{k^4}{\ell^2}\zeta - \frac{i}{\ell}\frac{(D-2)}{D}k^2 \mathcal{D}_\beta(\sqrt{\mathsf{h}}X^\beta) + \frac{i}{\ell}\sqrt{\mathsf{h}}k^ik^j(\partial_\beta h_{ij})X^\beta \\
  & = \frac{i}{\sqrt{h}}\frac{k^2}{\ell}\left(\pi^{bc}\delta \hat{h}_{bc} + p_A \delta\hat{\phi}^A + (\Pi^I)^{a_1 \ldots a_p}(\delta\hat{A}_I)_{a_1 \ldots a_p}\right)e^{ik\cdot x/\ell}
 \end{split}
\label{zeta equation}
\end{equation}

Where we have used that the induced metric on the tori ($h_{ij}$ in the above notation) is diagonal in these coordinates, as we are considering warped products of an asymptotically flat spacetime with a torus.

Note that the conformal factor $\psi$ does not appear in the linearised momentum constraints. However, it does appear in the linearised Hamiltonian constraint, which can be written as
\begin{equation}
\begin{split}
 0 &= (D-2)\mathcal{D}^\alpha\left(\sqrt{\mathsf{h}} \mathcal{D}_\alpha \psi\right) +\sqrt{\mathsf{h}}\left( V(\phi) - \frac{p}{2(p+1)!}G_{IJ}(\upd A^I)^{a_1 \ldots a_{p+1}}(\upd A^J)_{a_1 \ldots a_{p+1}} - \frac{D-2}{h}\pi^{ab}\pi_{ab} \right. \\
 & \left. - \frac{(D-2)}{2h}F^{AB}p_Ap_B - \frac{p!(D-p-2)}{2 h}G^{IJ}(\Pi_I)^{a_1 \ldots a_p}(\Pi_J)_{a_1 \ldots a_p} - (D-2)\frac{k^2}{\ell^2}\right)\psi + 2\sqrt{\frac{\mathsf{h}}{h}}\pi^{\alpha\beta}\mathcal{D}_\alpha X_\beta \\
 & + \sqrt{\frac{\mathsf{h}}{h}}\pi^{ij}(\partial_\alpha h_{ij})X^\alpha + \frac{2i}{\ell}\sqrt{\frac{\mathsf{h}}{h}}k^ik^j\pi_{ij}\zeta - \sqrt{\mathsf{h}}\frac{1}{\ell^2}k^ik^j\delta\hat{h}_{ij} - \frac{k^2}{\ell^2}\sqrt{\mathsf{h}}\delta\hat{h} 
\end{split}
\label{psi equation}
\end{equation}

Equations (\ref{X equation}), (\ref{zeta equation}) and (\ref{psi equation}) form a set of linear, uniformly elliptic\footnote{The principle symbol of $X^\alpha$ in (\ref{X equation}) is $\sigma(\xi)^a_{\phantom{a}b} = \xi^2 \delta^a_b + \frac{(D-2)}{D}\xi^a\xi_b$, which is invertible for all $\xi \neq 0$, and in fact $||\sigma(\xi)^a_{\phantom{a}b}|| \geq \xi^2$, where the operator norm is understood.} PDE for $X^\alpha$, $\zeta$ and $\psi$. In order to discuss their solutions, we must first find appropriate boundary conditions.

We take all members of our original family of spacetimes to obey our gauge conditions, and so the gauge conditions on our perturbations will be met as long as $\delta\vartheta = 0$. We find
\begin{equation}
 \delta\vartheta = \frac{(D-2)}{2}\mathscr{L}_\eta\psi + \frac{(D-1)}{2}H\psi - \frac{(D-2)}{D}\mathscr{L}_\eta(\eta_a X^a) + \frac{2}{D}H(\eta_a X^a) + \frac{2}{D}\mathscr{D}_a\left((h_{b}^{\phantom{b}a} - \eta_b\eta^a)X^b\right) + \frac{2}{D}\frac{i}{\ell}k^2\zeta
\end{equation}
where $\eta$ is the outward pointing normal to $B$ in $\Sigma$, $H$ is the trace of the extrinsic curvature of $B$ in $\Sigma$, and $\mathscr{D}$ is the induced covariant derivative on $B$. Since $B$ is the bifurcation surface of a black hole spacetime, we actually have $H = 0$, and so we will satisfy our gauge conditions if we impose mixed Neumann/Dirichlet boundary conditions at $B$ on our perturbations as follows: \begin{equation}
\begin{split}
 \mathscr{L}_\eta \psi\big|_B &= 0 \\
 \mathscr{L}_\eta (\eta_a X^a)\big|_B &= 0 \\
 (h_b^{\phantom{b}a} - \eta_b\eta^a)X^b\big|_B &= 0 \\
 \zeta\big|_B &= 0 \end{split}
\label{GM boundary}
\end{equation}
Existence of a solution to equations (\ref{X equation}) and (\ref{zeta equation}) with the above boundary conditions is guaranteed by the arguments given in appendix B of \cite{Hollands:2012sf}. In addition, the norm of this solution (in an appropriately weighted Sobolev space) is controlled by the norm of the right hand side of equation (\ref{zeta equation}), and tends to zero of order $\mathcal{O}(1/\ell)$ in the limit $\ell \rightarrow 0$.

Next we examine equation (\ref{psi equation}). We can multiply this by $\psi$ and integrate by parts, with the aim of showing that $||\psi||_{H^2} \rightarrow 0$ as $\ell \rightarrow \infty$, in $D \geq 6$. \footnote{The integrals do not all converge in $D = 4, 5$, but a limiting argument identical to that used in \cite{Hollands:2012sf} can be made to deal with these cases.} Note, however, that in order to make this argument, the second term in (\ref{psi equation}) must be negative, i.e.
\begin{equation}
\begin{split}
V(\phi) \leq &\frac{p}{2(p+1)!}G_{IJ}(\upd A^I)^{a_1 \ldots a_{p+1}}(\upd A^J)_{a_1 \ldots a_{p+1}} + \frac{D-2}{h}\pi^{ab}\pi_{ab} + \frac{(D-2)}{2h}F^{AB}p_Ap_B \\
& + \frac{p!(D-p-2)}{2 h}G^{IJ}(\Pi_I)^{a_1 \ldots a_p}(\Pi_J)_{a_1 \ldots a_p} + (D-2)\frac{k^2}{\ell^2}
\end{split}
\label{V condition}
\end{equation}
If this is not the case, then equation (\ref{psi equation}) tells us nothing about $||\psi||_{H^2}$. Note that the condition (\ref{V condition}) is a condition on the \emph{background} solution, and not on the perturbations: given a background solution, we could simply check whether this condition is satisfied or not. In particular, (\ref{V condition}) will be satisfied whenever $V = 0$, or when $\phi^A = 0$ in the background, i.e. when the scalar field potential vanishes, or when the scalar field itself vanishes in the background spacetime.

In the case where (\ref{V condition}) is not satisfied, it may be possible to make a more complicated ansatz for the perturbation (for example, making some additional perturbation to the scalar field and its momentum) in order to make the corresponding term negative. For example, when the scalar fields have vanishing canonical momenta, $p_A = 0$, we can eliminate the assumption (\ref{V condition}) if we make an additional perturbation
\begin{equation}
 \delta \phi^A = e^{ik\cdot x/\ell}(\delta \hat{\phi}^A - \psi\phi^A)
\end{equation}
where $\psi$ is the conformal factor introduced in (\ref{GM perts}). With this additional perturbation, the linearised momentum constraint is still satisfied. Under the additional assumptions $V_{,A}\phi^A - V \geq 0$ in the background, and $F_{A[B,C]} = 0$, we find a modified version of equation (\ref{psi equation}) of the form
\begin{equation}
\mathcal{D}^\alpha \left( f_1 \mathcal{D}_\alpha \psi\right) - f_2 \psi + f_3(X, \zeta, (1/\ell)\delta\hat{h}) = 0
\end{equation}
with $f_1$ and $f_2$ positive functions, and $f_3$ linear, and with appropriate decay properties for the approach used above to be applicable. In particular, $p_A = 0$ includes (but is not limited to) the case of a static black brane, for which all the canonical momenta vanish.

Finally, we note that the normalised canonical energy $\tilde{\mathcal{E}}(\delta S)$ differs from the normalised canonical energy of our original perturbations, $\tilde{\mathcal{E}}(\delta \hat{S})$, only by terms involving integrals of $X^\alpha$, $\zeta$, $\psi$ and their derivatives, all of which can be bounded by their norms in the above sense, and all of which are $\mathcal{O}(\ell^{-1})$. Note further that, since such quantities have been found to converge when their indices are contracted using the metric $h_{ab}$, they will certainly converge when contracted using quantities which decay faster near infinity, such as $\pi_{ab}$. Therefore, we conclude that \begin{equation}  \mathcal{E}_\ell(\delta S) = \mathcal{E}_\ell(\delta\hat{S}) + \mathcal{O}(\ell^{-1}) \end{equation} In particular, this means that we will have $\mathcal{E}_\ell(\delta S) < 0$ for $\ell$ sufficiently large, proving the Gubser-Mitra conjecture in this setting.

\section{Conclusion}

Hollands and Wald have previously constructed an energy functional for linearised perturbations in vacuum GR, whose positivity properties are connected with stability and instability, at least in some restricted class of perturbations. We have successfully generalised their construction to include a rather complicated matter model, and obtained the same results. Our model includes a large set of conserved charges, including magnetic and electric (possibly dipole) charges, all of which must be held constant in the class of perturbations under consideration. It is important to note that we did not include charged matter in our model - the ability to radiate charge would prevent us from arguing for instability.

In order to make our arguments we have had to investigate the stability of an appropriate notion of asymptotic flatness near null infinity in the presence of matter. We have found that an appropriate definition can be given, and that this definition is stable under linear perturbations, in the sense that linearised perturbations on some initial Cauchy surface, when evolved under the linearised equations of motion, will satisfy the linearised versions of our definitions of asymptotic flatness at null infinity. We should note that, for technical reasons, this only works in even dimensions $D >4$, though we expect that the case $D=4$ can be dealt with as a special case (as it is for the vacuum Einstein equations).

Our primary result is that an appropriate energy functional $\mathcal{E}$ can be defined on linearised perturbations to stationary solutions of our model. If we can prove that $\mathcal{E} > 0$ for all allowed perturbations, then the spacetime is stable to that class of linearised perturbations, whereas if $\mathcal{E} < 0$ for some perturbation, then the spacetime is unstable.

We have also extended the proof of the Gubser-Mitra conjecture given in \cite{Hollands:2012sf} to the case of charged black brane solutions to our model. The primary difficulty here is that the spacetimes under consideration are no longer simply Cartesian products, but this can be overcome by using suitable inner products (i.e. with weight $\sqrt{\mathsf{h}}$). We found that, for uniform, non-boosted black brane solutions to our Lagrangian, if the solution is both thermodynamically unstable and sufficiently extended, then it is classically unstable. Unfortunately, as in the vacuum case, our construction does not allow us to infer exactly how far the black brane must be extended before the instability will set in. In addition, we have had to place additional constraints on the scalar field potential in the background spacetime. In particular, we are able to deal completely with the case of a vanishing scalar field potential, or a vanishing scalar field in the background spacetime, or a static background spacetime. We expect an appropriate ansatz for the perturbations to yield the general case, perhaps with some further restrictions on the form of the potential $V(\theta)$ or the coupling terms $F_{AB}$ and $G_{AB}$.

The arguments given for stability and instability have a few shortcomings, however, the argument for instability whenever $\mathcal{E} < 0$ appears to be fairly strong. On the other hand, the primary criticism of the argument we have given for stability must be that we have restricted to axisymmetric perturbations. This unfortunate, especially when we reflect that a lot of interesting dynamics, such as the phenomena of superradiance (which is potentially problematic with regards to stability - see for example \cite{Hod:2009cp} and \cite{Shlapentokh-Rothman:2013ysa}), are lost when we restrict to axisymmetry. On the other hand, with some modification, it may be possible to adapt this method to deal with non-axisymmetric perturbations. Indeed, if our background spacetime has vanishing angular momenta, we can already drop this restriction.

\subsection*{Acknowledgements}
The author is very grateful to Harvey Reall for suggesting this project, as well as for many useful discussions and comments. This work was supported by the European Research Council Grant No. ERC-2011-StG 279363-HiDGR.

\pagebreak

\appendix
\section{Regular, Hyperbolic Equations Near Null Infinity}

Here we construct the hyperbolic PDE, which are the evolution equations for our rescaled variables presented in (\ref{null variables}), in our choice of gauge (\ref{asymp gauge}). First we consider the linearised, rescaled equations of motion for the matter fields. Note that these are all regular at $\Omega = 0$, which is the boundary of the conformally compactified manifold. See \cite{Geroch:1978ur}, \cite{Hollands:2003ie} for the vacuum case in $D = 4$ and $D \geq 6$ respectively. Note also that derivatives are taken with respect to the rescaled fields; for example 
\begin{equation}
 \tilde{V}_A = \frac{\partial}{\partial \varphi^A} V(\varphi) = \mathcal{O}(\Omega^{\frac{D}{2} - 1}) \end{equation}

In this section, we will make the same choice of conformal factor as in \cite{Hollands:2003ie}, which means that we can set $\Omega^{\frac{D}{2}}f = \tilde{n}_\mu \tilde{n}_\nu$ and $\Omega^(D/2 -2)\chi_{ab} = \tilde{\nabla}_a \tilde{n}_b$. We will also use the shorthand $H^I$ for the background field strength. Finally, for simplicity (of a sort) we will only treat the case where the gauge fields are one-forms in this section.

For completeness, we state here the conformal factors which relate our background and perturbed fields to their rescaled counterparts. The background fields are:
\begin{equation}
 \begin{split}
  g_{\mu\nu} &= \Omega^{-2}\tilde{g}_{\mu\nu} \\
  \phi^A &= \Omega^{\frac{D-2}{2}}\tilde{\phi}^A \\
  \upd A^I = H^I &= \Omega^{\frac{D-4-2p}{2}}\tilde{H}^I \\
  \imath_{\tilde{n}} \tilde{H}^I &= \Omega \hat{H}\\
  T_{\mu\nu} & = \Omega^{\frac{D-2}{2}}\tilde{T}_{\mu\nu}
 \end{split}
\end{equation}
Where $T_{\mu\nu}$ is the energy-momentum tensor.

The perturbations are:
\begin{equation}
 \begin{split}
  \delta g_{\mu\nu} &= \Omega^{\frac{D-6}{2}} \tau_{\mu\nu} \\
  \tilde{n}^\nu \tau_{\mu\nu} &= \Omega \hat{\tau}_\mu \\
  \tilde{g}^{\mu\nu} \tau_{\mu\nu} &= \Omega \tilde{\tau} \\
  \tilde{g}^{\mu\nu}\tilde{\nabla}_\mu \tau_\nu &= \sigma \\
  \delta\phi^A &= \Omega^{\frac{D-2}{2}}\delta\tilde{\phi}^A \\
  \delta A^I &= \Omega^{\frac{D-2-2p}{2}}\delta\tilde{A}^I \\
  \delta T_{\mu\nu} &= \Omega^{\frac{D-2}{2}}\delta{T}_{\mu\nu}
 \end{split}
\label{rescaled perturbation defs}
\end{equation}

Finally, we use the Lorentz gauge for the gauge field perturbations, $\text{div} \delta A^I = 0$, and we choose a modified transverse gauge for the metric,
\begin{equation}
 \nabla^\nu \delta g_{\nu\mu} - \frac{1}{2}\nabla_\mu \delta g - \Omega^{-1} n_\mu \delta g = 0
\end{equation}
which, in terms of the conformally rescaled quantities, takes the form
\begin{equation}
 \tilde{\nabla}^\nu \tau_{\nu\mu} - \frac{1}{2}\Omega\tilde{\nabla}_\mu \tilde{\tau} - \frac{D}{4}n_\mu \tau - \frac{D+2}{2} \hat{\tau}_\mu = 0
\label{rescaled gauge}
\end{equation}

Because of the excessive length of the equations in this section, we will only quote the important terms in the conformally rescaled equations of motion, which we present below. The terms which are omitted are at worst of order $\Omega^{\frac{D}{2} - 3}$, and they all contain at most one derivative of the perturbations.

\subsection{Matter Perturbation Equations of Motion}
The equations of motion for the scalar field perturbations are:
\begin{equation}
F_{AB}\tilde{\nabla}_\mu\tilde{\nabla}^\mu \delta\tilde{\phi}^B 
 + \Omega^{-1}\left(\frac{D}{2}-1\right)\left(\tilde{F}_{AB,C} + \tilde{F}_{AC,B} - \tilde{F}_{BC,A}\right)\tilde{\phi}^B \tilde{n}^\mu (\tilde{\nabla}_\mu \delta\tilde{\phi}^C )
 + \Omega^{-2}\tilde{V}_{AB}\delta\tilde{\phi}^B + \ldots = 0
\label{rescaled scalar field}
\end{equation}
Note that, in order for this equation to be regular, we require $\tilde{V}_{AB} = \mathcal{O}(\Omega^2)$ and $\tilde{F}_{AB,C} = \mathcal{O}(\Omega)$. This is an indication that the asymptotics of the scalar field change when the field is massive, or when the non-canonical kinetic term causes problems.

The equation of motion for the gauge field perturbations are
\begin{equation}
G_{IJ}\tilde{\nabla}_\nu\tilde{\nabla}^\nu \delta \tilde{A}^J_{\phantom{J} \mu} + \ldots = 0
\label{rescaled gauge field}
\end{equation}

And for the normal component of the perturbation to the gauge fields:
\begin{equation}
G_{IJ}\tilde{\nabla}_\mu\tilde{\nabla}^\mu \delta\hat{A}^J - \Omega^{-1}G_{IJ}\delta\tilde{A}^{J\mu}\tilde{n}^\nu \tilde{R}_{\mu\nu} + \ldots = 0
\label{normal rescaled gauge field}
\end{equation}
the second term in (\ref{normal rescaled gauge field}) appears divergent, however, though the use of the (conformally rescaled) Einstein equation, the Ricci tensor $\tilde{R}$ can be related to the rescaled energy momentum tensor, and so it can be seen to fall off fast enough. In fact, the contraction $\tilde{n}^\mu \tilde{R}_{\mu\nu}$ falls off even faster.

\subsection{Metric Perturbation Equations of Motion}
Next we examine the linearised and rescaled Einstein equation. The equation of motion for the metric perturbation is
\begin{equation}
 -\tilde{\nabla}_\rho \tilde{\nabla}^\rho \tau_{\mu\nu} + \frac{1}{2}\Omega \tilde{g}_{\mu\nu}\tilde{\nabla}_\rho\tilde{\nabla}^\rho\tilde{\tau} + \ldots = \delta \tilde{T}_{\mu\nu}
\label{rescaled metric perturbation}
\end{equation}

Note that there is a term involving second derivatives of the (rescaled) trace of the metric perturbation, which can be replaced by lower order terms by using the equation of motion for the trace of the metric perturbation, which is
\begin{equation}
\left(\frac{1}{2}D - 1\right)\tilde{\nabla}_\mu\tilde{\nabla}^\mu \tilde{\tau} + \ldots = \Omega^{-1}\tilde{g}^{\mu\nu}\delta\tilde{T}_{\mu\nu}
\label{trace rescaled metric perturbation}
\end{equation}
As before, the second term above is potentially problematic, but on closer inspection of the rescaled perturbation to the energy momentum tensor we find that its trace vanishes one power of $\Omega$ faster.

We also need equations of motion for the rescaled normal component of the metric perturbation, $\hat{\tau}_\mu$ and its divergence, $\sigma$. These are are found by taking the divergence of (\ref{rescaled metric perturbation}), and then taking the divergence once again, and using our gauge conditions.

The equation of motion for $\hat{\tau}_\mu$ is
\begin{equation}
-\frac{2-D}{D}\tilde{\nabla}_\nu \tilde{\nabla}^\nu \hat{\tau}_\mu + \frac{2-D}{D}\tilde{n}^\nu \tilde{\nabla}_\nu \tilde{\nabla}_\mu \tilde{\tau} + \ldots = \tilde{\nabla}^\nu\delta\tilde{T}_{\nu\mu}
\label{normal rescaled metric perturbation}
\end{equation}

The equation for $\sigma$ is
\begin{equation}
\begin{split}
&-\frac{(D-4)(D-2)}{2D} \left(\tilde{\nabla}_\mu\tilde{\nabla}^\mu\sigma - \Omega^{\frac{D}{2} - 2}\xi^{\mu\nu}\tilde{\nabla}_\rho\tilde{\nabla}^\rho\tau_{\mu\nu} + \Omega^{\frac{D}{2} - 1}\xi^{\mu\nu}\tilde{\nabla}_\mu\tilde{\nabla}_\nu\tilde{\tau} \right)
 + \frac{1}{2}\tilde{R}^{\mu\nu}\tilde{\nabla}_\rho \tilde{\nabla}^\rho \tau_{\mu\nu} + \ldots \\
& = \tilde{\nabla}^\mu\tilde{\nabla}^\nu \delta\tilde{T}_{\mu\nu}
\end{split}
\label{sigma equation}
\end{equation}

Note that, in both of the above equations, second derivatives of $\tilde{\tau}$ appear. Second derivatives of $\tau_{\mu\nu}$ also appear in (\ref{sigma equation}), however, we can use (\ref{rescaled metric perturbation}) to eliminate these in favour of lower order terms. Also, note that the coefficient of $\tilde{\nabla}_\mu\tilde{\nabla}^\mu \sigma$ in (\ref{sigma equation}) contains a factor of $(D-2)(D-4)$ - this same factor also appears in \cite{Hollands:2003ie}, and signifies that a different gauge or choice of variables is necessary in $D=4$.

One might worry that, since the (rescaled) perturbed energy-momentum tensor, $\delta\tilde{T}_{\mu\nu}$, already contains first derivatives of the perturbations to the matter fields, the above equations might contain third and fourth derivatives. However, since $\delta \left(\nabla_\mu T^{\mu\nu}\right) = 0$,we can write the right-hand sides of (\ref{normal rescaled metric perturbation}) and (\ref{sigma equation}) as
\begin{equation}
 \begin{split}
 \tilde{\nabla}^\nu \delta\tilde{T}_{\mu\nu} &=
 - \Omega^{\frac{D}{2} - 1}\tilde{T}^{\nu\rho}\tilde{\nabla}_\mu(\tau_{\nu\rho})
 - \Omega^{\frac{D}{2} - 1}\tau^{\nu\rho}\tilde{\nabla}_\rho(\tilde{T}_{\mu\nu})
 - \frac{1}{2}D\tilde{n}_\mu\Omega^{\frac{D}{2} - 2}\tau^{\nu\rho}\tilde{T}_{\nu\rho}
 - \frac{1}{2}D\tilde{n}^\nu\Omega^{\frac{D}{2} - 2}\tilde{T}_{\mu\rho}\tau_\nu^{\phantom{\nu}\rho} \\
 &+ n^\nu\Omega^{\frac{D}{2} - 2}\tilde{T}_{\mu\rho}\tau_\nu^{\phantom{\nu}\rho}
 - n^\nu\Omega^{\frac{D}{2} - 2}\tau_\mu^{\phantom{\mu}\rho}\tilde{T}_{\nu\rho}
 - \frac{D\tilde{n}^\nu\delta\tilde{T}_{\mu\nu}}{2\Omega}
 + \frac{n_\mu\delta\tilde{T}^\nu_{\phantom{\nu}\nu}}{\Omega}
 + \frac{n^\nu\delta\tilde{T}_{\mu\nu}}{\Omega}
 - \tilde{\tau}\tilde{n}^\nu\Omega^{\frac{D}{2} - 1}\tilde{T}_{\mu\nu} \\
 &- \hat{\tau}^\nu\Omega^{\frac{D}{2} - 1}\tilde{T}_{\mu\nu}
 \end{split}
\label{rescaled energy momentum 1}
\end{equation}
and
\begin{equation}
 \begin{split}
 \tilde{\nabla}^\mu \tilde{\nabla}^\nu \delta\tilde{T}_{\mu\nu} &=
 - \frac{1}{2}D\tilde{n}^\mu\Omega^{\frac{D}{2} - 2}\tilde{T}^{\nu\rho}\tilde{\nabla}_\mu(\tau_{\nu\rho})
 - \frac{1}{2}D\tilde{n}^\mu\Omega^{\frac{D}{2} - 2}\tau^{\nu\rho}\tilde{\nabla}_\mu(\tilde{T}_{\nu\rho})
 - \frac{1}{2}D\tilde{n}^\mu\Omega^{\frac{D}{2} - 2}\tilde{T}^{\nu\rho}\tilde{\nabla}_\rho(\tau_{\mu\nu}) \\
 &+ n^\mu\Omega^{\frac{D}{2} - 2}\tilde{T}^{\nu\rho}\tilde{\nabla}_\rho(\tau_{\mu\nu})
 - n^\mu\Omega^{\frac{D}{2} - 2}\tilde{T}_\mu^{\phantom{\mu}\nu}\tilde{\nabla}_\rho(\tau_\nu^{\phantom{\nu}\rho})
 - n^\mu\Omega^{\frac{D}{2} - 2}\tau^{\nu\rho}\tilde{\nabla}_\rho(\tilde{T}_{\mu\nu}) \\
 &- \Omega^{\frac{D}{2} - 1}\tau^{\mu\nu}\tilde{\nabla}_\rho(\tilde{\nabla}_\nu(\tilde{T}_\mu^{\phantom{\mu}\rho}))
 - \Omega^{\frac{D}{2} - 1}\tilde{T}^{\mu\nu}\tilde{\nabla}_\rho(\tilde{\nabla}^\rho(\tau_{\mu\nu}))
 - \Omega^{\frac{D}{2} - 1}\tilde{\nabla}^\rho(\tau^{\mu\nu})\tilde{\nabla}_\nu(\tilde{T}_{\mu\rho}) \\
 &- \Omega^{\frac{D}{2} - 1}\tilde{\nabla}^\rho(\tau^{\mu\nu})\tilde{\nabla}_\rho(\tilde{T}_{\mu\nu})
 + \frac{n^\mu\tilde{\nabla}_\mu(\delta\tilde{T}^\nu_{\phantom{\nu}\nu})}{\Omega}
 - \tilde{\tau}\tilde{n}^\mu\Omega^{\frac{D}{2} - 1}\tilde{\nabla}_\nu(\tilde{T}_\mu^{\phantom{\mu}\nu})
 - n^\mu\Omega^{\frac{D}{2} - 1}\tilde{T}_\mu^{\phantom{\mu}\nu}\tilde{\nabla}_\nu(\tilde{\tau}) \\
 &- \frac{1}{2}D\hat{\tau}^\mu\Omega^{\frac{D}{2} - 1}\tilde{\nabla}_\nu(\tilde{T}_\mu^{\phantom{\mu}\nu})
 - \Omega^{\frac{D}{2} - 1}\tilde{T}^{\mu\nu}\tilde{\nabla}_\nu(\hat{\tau}_\mu)
 - \frac{1}{2}D\Omega^{D  - 3}\tau^{\mu\nu}\tilde{T}_\mu^{\phantom{\mu}\rho}\chi_{\nu\rho}
 - \frac{1}{2}D\Omega^{D  - 3}\chi^\rho_{\phantom{\rho}\rho}\tau^{\mu\nu}\tilde{T}_{\mu\nu} \\
 &+ \frac{1}{2}D\tilde{n}^\mu\hat{\tau}^\nu\Omega^{\frac{D}{2} -  2}\tilde{T}_\mu^{\phantom{\mu}\rho}\tilde{g}_{\nu\rho}
 + \frac{D^2\tilde{n}^\mu\tilde{n}^\nu\delta\tilde{T}_{\mu\nu}}{4\Omega^2}
 - \frac{D\tilde{n}^\mu\tilde{n}^\nu\delta\tilde{T}_{\mu\nu}}{2\Omega^2}
 - \frac{1}{2}D\Omega^{\frac{D}{2} - 2}\delta\tilde{T}^{\mu\nu}\chi_{\mu\nu} \\
 &+ \Omega^{\frac{D}{2} - 2}\delta\tilde{T}^{\mu\nu}\chi_{\mu\nu}
 + \Omega^{\frac{D}{2} - 2}\delta\tilde{T}^\mu_{\phantom{\mu}\mu}\chi^\nu_{\phantom{\nu}\nu}
 + \frac{1}{2}Df\Omega^{D  - 3}\tau^{\mu\nu}\tilde{T}_{\mu\nu}
 - \tilde{\tau}\Omega^{D  - 2}\tilde{T}^{\mu\nu}\chi_{\mu\nu}
 - \frac{1}{2}Df\Omega^{\frac{D}{2} - 2}\delta\tilde{T}^\mu_{\phantom{\mu}\mu}
 \end{split}
\label{rescaled energy momentum 2}
\end{equation}

We see that, as long as $\delta \tilde{T}_{\mu\nu} = \mathcal{O}(\Omega^2)$, these terms are regular at $\Omega = 0$. As it turns out, $\delta T_{\mu\nu} = \mathcal{O}(\Omega^{\frac{D}{2} - 3})$, so we are in good shape if $D \geq 10$. In fact, a more careful analysis of the terms which appear to diverge in (\ref{rescaled energy momentum 1}) and (\ref{rescaled energy momentum 2}) reveals that we only need $D \geq 6$.

\subsection{Hyperbolicity}
Thus, for $D \geq 6$, the set of equations (\ref{rescaled scalar field}) through (\ref{sigma equation}) form a closed set of regular, second order PDE. We now claim that this system is also hyperbolic. For a system of PDE, of the form
\begin{equation}
 P_{XY}^{\mu\nu} \nabla_\mu \nabla_\nu u^Y + (\text{lower order terms}) = 0
\end{equation}
the principle symbol is defined via the determinant of the matrix $P$, taken over its $X,Y$ indices, i.e.
\begin{equation}
 P(\xi) = \det(P_{XY}^{\mu\nu} \xi_\mu \xi_\nu)
\end{equation}
Because of the form of equations (\ref{rescaled scalar field})-(\ref{sigma equation}), we can take the matrix $P_{XY}^{\mu\nu} \xi_\mu \xi_\nu$ to be upper-triangular, with diagonal entries $\xi^2$. Thus this system of equations is hyperbolic.

\subsection{Gauge Fixing}
There is one final issue which we need to deal with in this appendix: the gauge fixing. In particular, the gauge we fix for the matter perturbations may potentially cause problems, since it involves a negative power of $\Omega$ when expressed in terms of the non-rescaled quantities.

If we make a change of gauge which respects our asymptotic flatness conditions, then we are making the change
\begin{equation}
 \delta g_{\mu\nu} \rightarrow \delta g_{\mu\nu} + \nabla_\mu X_\nu + \nabla_\nu X_\mu
\end{equation}
Where we can conformally rescale the gauge vector $X_\mu$ according to
\begin{equation}
 X^\mu = \Omega^{\frac{D-4}{2}} \tilde{X}^\mu
\end{equation}
In order to obey our asymptotic flatness conditions, the component of $X^\mu$ in the $\tilde{n}$ direction must fall off one power of $\Omega$ faster, and so we can define
\begin{equation}
 \tilde{n}^\mu \tilde{X}_\mu = \Omega \eta
\end{equation}
Then, in terms of conformally rescaled variables, our gauge conditions become
\begin{equation}
\begin{split}
 &\tilde{g}^{\nu\rho}\tilde{\nabla}_\rho(\tilde{\nabla}_\nu(\tilde{X}_\mu))
 + \frac{\tilde{\nabla}_\nu(\tau_\mu^{\phantom{\mu}\nu})}{\Omega}
 + D\tilde{X}^\nu\Omega^{\frac{D}{2} - 2}\chi_{\mu\nu}
 + \frac{1}{2}D\tilde{X}_\mu\Omega^{\frac{D}{2} - 2}\chi^\nu_{\phantom{\nu}\nu}
 + \tilde{X}^\nu\tilde{R}_{\mu\nu}
 - 2\tilde{\nabla}_\mu(\eta)
 - \frac{1}{2}\tilde{\nabla}_\mu(\tilde{\tau}) \\
 &- \frac{1}{4}D^2f\tilde{X}_\mu\Omega^{\frac{D}{2} - 2}
 - \frac{1}{2}Df\tilde{X}_\mu\Omega^{\frac{D}{2} - 2}
 - \frac{D\tilde{\tau}\tilde{n}_\mu}{4\Omega}
 - \frac{D\hat{\tau}_\mu}{2\Omega}
 - \frac{\hat{\tau}_\mu}{\Omega} = 0
\end{split}
\label{gauge vector}
\end{equation}
and
\begin{equation}
\begin{split}
 &\frac{n^\mu\tilde{\nabla}_\nu(\tau_\mu^{\phantom{\mu}\nu})}{\Omega}
 - 2\Omega^{\frac{D}{2} - 1}\chi^{\mu\nu}\tilde{\nabla}_\nu(\tilde{X}_\mu)
 - \tilde{X}^\mu\Omega^{\frac{D}{2} - 1}\tilde{\nabla}_\nu(\chi_\mu^{\phantom{\mu}\nu})
 + \Omega\tilde{g}^{\mu\nu}\tilde{\nabla}_\nu(\tilde{\nabla}_\mu(\eta))
 + \frac{1}{2}D\tilde{X}^\mu\tilde{n}^\nu\Omega^{\frac{D}{2} - 2}\chi_{\mu\nu} \\
 &+ \tilde{X}^\mu\tilde{n}^\nu\Omega^{\frac{D}{2} - 2}\chi_{\mu\nu}
 + \tilde{X}^\mu\tilde{n}^\nu\tilde{R}_{\mu\nu}
 + \frac{1}{2}D\eta\Omega^{\frac{D}{2} - 1}\chi^\mu_{\phantom{\mu}\mu}
 + \eta\Omega^{\frac{D}{2} - 1}\chi^\mu_{\phantom{\mu}\mu}
 - \frac{1}{2}\tilde{n}^\mu\tilde{\nabla}_\mu(\tilde{\tau})
 - \frac{D\tilde{n}^\mu\hat{\tau}_\mu}{2\Omega}
 - \frac{n^\mu\hat{\tau}_\mu}{\Omega} \\
 &- \frac{1}{4}D^2\eta f\Omega^{\frac{D}{2} - 1}
 - \frac{1}{2}D\eta f\Omega^{\frac{D}{2} - 1}
 - \frac{1}{4}Df\tilde{\tau}\Omega^{\frac{D}{2} - 1} = 0
\end{split}
\label{normal gauge vector}
\end{equation}

Note that both of the above equations are regular at $\Omega = 0$ as long as
\begin{equation}
 \tilde{\nabla}^\nu \tau_{\nu\mu} - \frac{D+2}{2}\tau_\mu - \frac{D}{4}\tau n_\mu = \mathcal{O}(\Omega)
\label{bad terms}
\end{equation}
However, an arbitrary solution to the equations of motion for the rescaled metric and matter fields will not satisfy (\ref{bad terms}), unless we perform a change of gauge, which involves solving (\ref{gauge vector}) and (\ref{normal gauge vector}). The source terms in these equations diverge at $\Omega = 0$, which may seem like bad news. However, we do not require any kind of control over the size of the gauge vector field $X^\mu$ near infinity, and so merely establishing existence of a solution for $\Omega > 0$ should suffice.

A solution to equations (\ref{gauge vector}) and (\ref{normal gauge vector}) in the region $\Omega > 0$ exists by the following argument: let $\epsilon > 0$. Then a solution to (\ref{gauge vector}) and (\ref{normal gauge vector}) exists in the region $\Omega > \epsilon$, since we can replace $\Omega^{-1}$ in equations (\ref{gauge vector}) and (\ref{normal gauge vector}) with
\begin{equation}
 \bar{\Omega} = \begin{cases} \Omega &\mbox{if } \Omega > \epsilon \\
    \text{smooth} &\mbox{if } \epsilon \geq \Omega \geq 0
                \end{cases}
\end{equation}
A solution to this equation exists by the previous arguments, and it will also solve the original equations in the region $\Omega > 0$, by the ``domain of dependence property'' of the wave equation i.e. solutions in some region depend only on the data and equations of motion in the causal past of that region. Of course, this does not mean that these solutions will have a well-defined limit as $\Omega \rightarrow 0$, but we do not require that our gauge vector is well defined in this limit. We also note that the standard transverse gauge, after conformal rescaling, becomes
\begin{equation}
 \tilde{\nabla}^\nu \tau_{\nu\mu} - \frac{1}{2}\Omega\tilde{\nabla}_\mu \tilde{\tau} + \left(1 - \frac{D}{4}\right)n_\mu \tau - \frac{D+2}{2} \hat{\tau}_\mu = 0
\end{equation}
This is almost identical to (\ref{rescaled gauge}), and so imposing it engenders the same issues.

Another interesting observation is the following: suppose we have a solution to the equations of motion in some other gauge, in which the perturbations are still asymptotically flat (meaning that the variables defined in (\ref{rescaled perturbation defs}) are smooth up to the boundary of the manifold). Then we claim that the necessary condition for our desired gauge change, (\ref{bad terms}), is in fact satisfied. The reason for this is that these terms, multiplied by $\Omega^{-1}$, are precisely the apparently divergent terms which appear in the equations of motion for the conformally rescaled metric (this was the reason for our choice of gauge). Thus, if we already have a solution to these equations of motion, (\ref{bad terms}) must be satisfied, since we can relate these terms to regular ones through the equations of motion.

\section{Hamiltonian Formalism}

In this section we recast the field equations into Hamiltonian form (see \cite{Arnowitt:1962hi}, \cite{Arnowitt:1960es}). Choosing a spacelike hypersurface $\Sigma$ with unit normal $n^\mu$, and a timelike vector $t^\mu$ we perform the usual decomposition of $t^\mu$ into the lapse $N$ and shift $N^a$, where from now on Latin indices will refer to tensors on $\Sigma$ and Greek indices to tensors $\mathcal{M}$.

The fields appearing in the Hamiltonian will be the restriction to $\Sigma$ of the metric, $h_{ab}$ and the momentum canonically conjugate to this, $\pi^{ab}$; the scalar fields $\phi^A$ and their canonical momenta $p_A$; the restriction to $\Sigma$ of the gauge fields $(A^I)_{a \ldots}$ and the momenta canonically conjugate to these, $(\Pi_I)^{a \ldots}$. The Hamiltonian also includes some unphysical fields representing the gauge degrees of freedom: the lapse function $N$, the shift vector $N^a$, and the normal components of the gauge fields $\mathcal{V}^I = \imath_n A^I$. With these definitions, the Hamiltonian is found to be
\begin{equation}
 H=\int_\Sigma \left(N \mathcal{C}_0 + N^a\mathcal{C}_a + N(\mathcal{V}^I)^{a_1 \ldots a_{p-1}} (\mathcal{C}_{(\mathcal{V}^I)})_{a_1 \ldots a_{p-1}}\right)
\end{equation}
where $C_0 = 0$ is the Hamiltonian constraint, $C_a=0$ is the momentum constraint and $C_{\mathcal{V}^I}=0$ are the Gauss law constraints, and surface terms have been neglected. The constraints are given explicitly by:
\begin{equation}
 \begin{split}
  \mathcal{C}_0 & = -\sqrt{h}R + \frac{1}{\sqrt{h}}\left(\pi^{ab}\pi_{ab} - \frac{1}{D-2}\pi^2\right) + \frac{1}{2}\sqrt{h}F_{AB}\partial_a\phi^A \partial^a \phi^B + \frac{1}{2}\frac{1}{\sqrt{h}}F^{AB}p_Ap_B\\
  & \phantom{AAA}+ \sqrt{h}V(\phi) + \frac{1}{2(p+1)!}\sqrt{h}G_{IJ}(\upd A^I)^{a_1 \ldots a_{p+1}}(\upd A^J)_{a_1 \ldots a_{p+1}} + \frac{p!}{2}\frac{1}{\sqrt{h}}G^{IJ}(\Pi_I)^{a_1 \ldots a_p} (\Pi_J)_{a_1 \ldots a_p} \\
  \mathcal{C}_a & = -2\sqrt{h}D_b\left(\frac{\pi_a^{\phantom{a}b}}{\sqrt{h}}\right) + p_A\partial_a\phi^A + (p+1)\partial_{[a}A^I_{\phantom{I}b_1 \ldots b_p]}\Pi_I^{\phantom{I}b_1 \ldots b_p} \\
  \mathcal{C}_{(\mathcal{V}^I)} & = (-1)^{(D-1)p + D}\sqrt{h}*\upd *\left(\Pi^\flat_{I}\sqrt{h}\right) \\
 \end{split}
\end{equation}
where $\Pi_I^\flat$ is the $(n-1)$-form density formed by lowering the indices on $\Pi_I$ with $h_{ab}$.

We also require the linearised versions of the constraint equations. Satisfaction of these (when linearised about a solution to the constraint equations) is equivalent to:
\begin{equation}
 \begin{split}
  0 &= R^{ab}\delta h_{ab} - D^aD^b\delta h_{ab} + D^c D_c \delta h - \frac{1}{h}\left(\pi^{ab}\pi_{ab} - \frac{1}{D-2}\pi^2\right)\delta h + \frac{2}{h}\pi^{ca}\pi^b_{\phantom{b}c}\delta h_{ab} - \frac{2}{D-2}\frac{1}{h}\pi\pi^{ab}\delta h_{ab} \\
  & \phantom{AAA}- \frac{1}{2}F_{AB}\partial^a\phi^A\partial^b\phi^B\delta h_{ab} - \frac{1}{2}F^{AB}\frac{1}{h} p_Ap_B \delta h - \frac{1}{2p!}G_{IJ}(\upd A^I)^{ac_1 \ldots c_p}(\upd A^J)^b_{\phantom{b}c_1 \ldots c_p}\delta h_{ab} \\
  & \phantom{AAA}- \frac{1}{2}G^{IJ}(\Pi_I)^{a_1 \ldots a_p}(\Pi_J)_{a_1 \ldots a_p} \delta h + \frac{pp!}{2}\frac{1}{h}G^{IJ}(\Pi_I)^{ac_1 \ldots c_{p-1}}(\Pi_J)^b_{\phantom{b}c_1 \ldots c_{p-1}}\delta h_{ab} + \frac{2}{h}\pi_{ab}\delta \pi^{ab} \\
  & \phantom{AAA} - \frac{2}{h}\frac{1}{D-2}\pi\delta\pi + F_{AB}\partial^a\phi^A\partial_a\delta\phi^B + \frac{1}{2(p+1)!}G_{IJ,A}(\upd A^I)^{a_1 \ldots a_{p+1}}(\upd A^J)_{a_1 \ldots a_{p+1}} \delta \phi^A\\
  & \phantom{AAA} + \frac{p!}{2}G^{IJ}_{\phantom{IJ},A}(\Pi_I)^{a_1 \ldots a_p}(\Pi_J)_{a_1 \ldots a_p} \delta \phi^A + V_{,A}\delta \phi^A+ \frac{1}{2}F_{AB,C}(\partial_a \phi^A \partial^a \phi^B)\delta\phi^C + \frac{1}{2}\frac{1}{h}F^{AB}_{\phantom{AB},C}p_Ap_B\delta\phi^C\\
  & \phantom{AAA} + \frac{1}{2}\frac{1}{h} F^{AB}p_A \delta p_B + G_{IJ}(\upd A^I)^{a_1 \ldots a_{p+1}}(\upd \delta A^J)_{a_1 \ldots a_{p+1}} + p!G^{IJ}(\Pi_I)^{a_1 \ldots a_p}(\delta \Pi_J)_{a_1 \ldots a_p} \\
  0 &= -2\sqrt{h}D_c\left(\frac{\pi^{cb}}{\sqrt{h}}\right)\delta h_{ab} - 2\pi^{bc}D_c\delta h_{ab} + \pi^{bc}D_a\delta h_{bc} - 2\sqrt{h}D_b\left(\frac{\delta\pi_a^{\phantom{a}b}}{\sqrt{h}}\right) + p_A\partial_a\delta\phi^A + \delta p_A \partial_a \phi^A \\
  & \phantom{AAA} + (p+1)\partial_{[a}(\delta A_I)_{b_1 \ldots b_p]}(\Pi^I)^{b_1 \ldots b_p} + (p+1) \partial_{[a}(A^I)_{b_1 \ldots b_p]} (\delta \Pi_I)^{b_1 \ldots b_p} \\
  0 &= \upd * \left(\frac{\delta\Pi_I^\flat}{\sqrt{h}}\right) \\
 \end{split}
\label{linear}
\end{equation}

\end{document}